\documentclass[a4paper,leqno]{amsart}

\usepackage{latexsym}
\usepackage[english]{babel}
\usepackage{fancyhdr}
\usepackage[mathscr]{eucal}
\usepackage{amsmath}
\usepackage{mathrsfs}
\usepackage{amsthm}
\usepackage{amsfonts}
\usepackage{amssymb}
\usepackage{amscd}
\usepackage{bbm}
\usepackage{graphicx}
\usepackage{subcaption}
\usepackage{graphics}
\usepackage{latexsym}
\usepackage{color}

\usepackage{pifont}
\usepackage{booktabs} 

\newcommand{\ud}{\mathrm{d}}

\newcommand{\ii}{\mathrm{i}}
\newcommand{\cH}{\mathcal{H}}

\theoremstyle{plain}
\newtheorem{theorem}{Theorem}[section]

\theoremstyle{definition}

\numberwithin{equation}{section}

\begin{document}

\title[LR bounds and growth of correlations in Bose mixtures]
{Lieb-Robinson bounds and growth of correlations in Bose mixtures}

\author[Alessandro Michelangeli]{Alessandro Michelangeli}
\address[A.~Michelangeli]{Institute for Applied Mathematics, and Hausdorff Center for Mathematics, University of Bonn \\ Endenicher Allee 60 \\ 
D-53115 Bonn (GERMANY).}
\email{michelangeli@iam.uni-bonn.de}

\author[Nicola Santamaria]{Nicola Santamaria}
\address[N.~Santamaria]{Department of Mathematics, University of Rome La Sapienza \\ Piazzale Aldo Moro 5 \\ 
00185 Rome (ITALY).}
\email{nicola.santamaria@uniroma1.it}


\begin{abstract}
 For a mixture of interacting Bose gases initially prepared in a regime of condensation (uncorrelation), it is proved that in the course of the the time evolution observables of disjoint sets of particles of each species have correlation functions that remain asymptotically small in the total number of particles and display a controlled growth in time. This is obtained by means of ad hoc estimates of Lieb-Robinson type on the propagation of the interaction, established here for the multi-component Bose mixture. 
\end{abstract}

\date{\today}

\subjclass[2020]{}
\keywords{Bose mixtures, composite BEC, many-body dynamics, correlation function, Lieb-Robinson bounds
}

\thanks{This work is partially supported by the Alexander von Humboldt Foundation. We are also grateful to P.~T.~Nam for valuable discussions and for the kind hospitality that he gave to one of us (N.S.) in the course of the preparation of this work, as well as to F.~Benatti and J.~Lee for further enlightening exchanges and feedback on the whole subject.}

\maketitle


\section{Set-up and main results}\label{intro}

 In this note we demonstrate a typical behaviour of the quantum dynamics of mixtures of different species of identical bosonic particles with non-trivial interaction (henceforth also \emph{interacting Bose mixtures}), asymptotically in the total number $N$ of particles, which generalises the analogous feature for single-species Bose gases.

 It consists of an asymptotic-in-$N$ vanishing of correlation functions between observables involving different particles of the same species and of distinct species, and a controlled growth in time, in the course of the time evolution when the system is initially prepared in an uncorrelated (`condensate') state.

 The main analytic tool to obtain such description, which is a result of relevance per se, is a bound of Lieb-Robinson type quantifying the increasing lack of commutativity of the above-mentioned observables that commute instead at time zero.

 This also provides the mathematical apparatus for deriving from the many-body quantum (Schr\"{o}dinger) evolution, in the limit $N\to\infty$, the system of coupled non-linear Schr\"{o}dinger equations describing the effective dynamics of the Bose mixture, a point of view that we will not push further, however, since by now the same effective equations have been derived through many other and often much more versatile methods.

 The mathematical description of interacting Bose mixtures and their time evolution is directly motivated by the latest advances in experiments with cold atoms and Bose-Einstein condensation (BEC) \cite{pethick02,pita-stringa-2016}. The confinement and cooling of Bose mixtures at ultra-low temperature and their observation in time have been actually on top of the experimental agenda since the very first realisation of BEC: they can be prepared as atomic gases of the same element, typically $^{87}\mathrm{Rb}$, which occupy two hyperfine states with no interconversion between particles of different hyperfine states \cite{MBGCW-1997,Matthews_HJEWC_DMStringari_PRL1998,HMEWC-1998,Hall-Matthews-Wieman-Cornell_PRL81-1543}, or also as heteronuclear mixtures such as $^{41}\mathrm{K}$-$^{87}\mathrm{Rb}$ \cite{Modugno-Ferrari-Inguscio-etal-Science2001_multicompBEC}, $^{41}\mathrm{K}$-$^{85}\mathrm{Rb}$ \cite{Modugno-PRL-2002}, $^{39}\mathrm{K}$-$^{85}\mathrm{Rb}$ \cite{MTCBM-PRL2004_BEC_heteronuclear}, and $^{85}\mathrm{Rb}$-$^{87}\mathrm{Rb}$ \cite{Papp-Wieman_PRL2006_heteronuclear_RbRb}. 
	A comprehensive review of the related physical properties may be found in \cite[Chapter 21]{pita-stringa-2016}.

 Mathematically, the subject has recently attracted attention in the framework of the study of the ground state properties of Bose gases and of the rigorous derivation of effective dynamical equations from the many-body Schr\"{o}dinger evolution \cite{LSeSY-ober,Benedikter-Porta-Schlein-2015}, an already flourishing subject for \emph{single-species bosonic systems} with a wide spectrum of results for various spatial dimensions $d =1, 2, 3$, in different scaling limits in $N$, with various local singularities of the interaction potentials, and through a multitude of different techniques
(kinetic-hierarchical methods, projection-counting methods, Fock space methods, semi-classical
measure-theoretic methods, and so on and so forth).

  The rigorous analysis for \emph{interacting Bose mixtures} has been developed in \cite{M-Olg-2016_2mixtureMF,AO-GPmixture-2016volume,Anap-Hott-Hundertmark-2017,DeOliveira-Michelangeli-2016,MNO-2017,Michelangeli-Pitton-2018,LeeJ-mixt2020-JMP2021}. Other types of \emph{composite} condensation (meaning, BEC with some sort of internal structure) analogous to condensate mixtures have been analysed mathematically in the case of \emph{pseudo-spinor condensates} \cite{MO-pseudospinors-2017}, \emph{spinor condensates} \cite{MO-2018-spin-spin}, and \emph{fragmented condensates} \cite{DimFalcOlg21-fragmented}.

 Let us introduce the concrete set-up for the present analysis. A system of $N_1+N_2$ non-relativistic quantum particles in $d$ spatial dimensions is considered, consisting respectively of $N_1$ and $N_2$ identical particles of two distinct bosonic species, say, A and B, coupled by both inter-species and intra-species two-body interactions. As is typical in a variety of experiments, such interactions are assumed to depend, isotropically, on the particle distance only, and do not couple spins; in particular, no intra-species transition is allowed. The natural mathematical setting to describe the states of such a Bose mixture, in a formalism of zero temperature, is therefore the Hilbert space
 \begin{equation}\label{eq:H12-tensorproduct}
  \cH_{N_1,N_2}\;:=\;L^2(\mathbb{R}^{N_1 d},\ud x_1\cdots\ud x_{N_1})\otimes L^2(\mathbb{R}^{N_2 d},\ud y_1\cdots\ud y_{N_1})\,,
 \end{equation}
 and more precisely the states of the system can only occupy the Bosonic sector of $\cH_{N_1,N_2}$, namely the Hilbert subspace 
  \begin{equation}
  \cH_{N_1,N_2,\mathrm{sym}}\;:=\;L^2_{\mathrm{sym}}(\mathbb{R}^{N_1 d},\ud x_1\cdots\ud x_{N_1})\otimes L^2_{\mathrm{sym}}(\mathbb{R}^{N_2 d},\ud y_1\cdots\ud y_{N_1})\,,
 \end{equation}
 thus consisting of square-integrable functions 
 \[
  \Psi(x_1,\dots,x_{N_1};y_1,\dots,y_{N_2})
 \]
 of two sets of variables, which are invariant under exchange of any two $x$-variables or any two $y$-variables (with no apriori overall permutation symmetry among the two sets of variables). The whole present analysis will apply more generally to an arbitrary number of components.

 At an ample level of generality, the Hamiltonian for the above mixture, namely the generator of the time evolution, is a self-adjoint realisation of an operator of the form
\begin{equation}\label{eq:HN1N2}
\begin{split}
H_{N_1,N_2}\;=\;&\sum_{i=1}^{N_1}(h_1)_i^\mathrm{A}+\frac{1}{N_1}\sum_{i<j}^{N_1}V_1(x_i-x_j) \\
+\;&\sum_{r=1}^{N_2}(h_2)_r^\mathrm{B}+\frac{1}{N_2}\sum_{r<s}^{N_2}V_2(y_r-y_s) \\
&\quad +\frac{1}{N_1+N_2}\sum_{i=1}^{N_1}\sum_{r=1}^{N_2}V_{12}(x_i-y_r)
\end{split}
\end{equation}
 acting on $\cH_{N_1,N_2,\mathrm{sym}}$.

 The various terms and the notation in \eqref{eq:HN1N2} are understood as usual as follows \cite{M-Olg-2016_2mixtureMF}. $h_1$ and $h_2$ are (possibly different) one-body Schr\"{o}dinger operators acting 
 self-adjointly on $L^2(\mathbb{R}^d)$, of the form
 \begin{equation}\label{eq:h-op}
  h\;=\;-\Delta_x+U(x)
 \end{equation}
 in suitable units, where $U$ is a real-valued measurable function on $\mathbb{R}^d$ modelling a spatial confinement (in practice, a harmonic potential or a potential well). Other variants of \eqref{eq:h-op} may be accommodated in the present analysis, such as a Schr\"{o}dinger operator with external magnetic field, or also its semi-relativistic version. By $(h_1)_i^\mathrm{A}$ one means the operator on $\cH_{N_1,N_2}$ acting as $h_1$ on the $i$-th variable of A-type, thus on $x_i$, and trivially as the identity on all variables, and analogous meaning is attributed to $(h_2)_i^\mathrm{B}$. Each of the $(h_1)_i^\mathrm{A}$'s or $(h_2)_i^\mathrm{B}$'s separately does not keep $\cH_{N_1,N_2,\mathrm{sym}}$, but of course their sums in \eqref{eq:HN1N2} do. The $V_\alpha$'s, with the label $\alpha$ running on $\{1,2,12\}$, are measurable $\mathbb{R}^d\to\mathbb{R}$ functions -- in physical applications they are typically invariant under rotations, modelling a two-body spatially isotropic interaction, but such an assumption is not needed here.

 In particular, $V_1$ and $V_2$ model the \emph{intra-species} coupling, whereas $V_{12}$ describes the interaction \emph{between different species}, and if $V_{12}\equiv 0$, then the Hamiltonian $H_{N_1,N_2}$ would be exactly decoupled as the product of two independent Hamiltonian, one for each species (thus, acting non-trivially only on one factor of the tensor product in \eqref{eq:H12-tensorproduct}).

 The numerical pre-factors in front of each $V_\alpha$ are chosen so as to prepare the analysis of the limit $N_1,N_2\to\infty$ in the \emph{mean-field scaling limit} \cite{am_GPlim}. One should therefore understand that for the actual numbers $N_1^{\mathrm{exp}}$ and $N_2^{\mathrm{exp}}$ of particles in a concrete experiment, $(N_1^{\mathrm{exp}})^{-1}V_1$, $(N_2^{\mathrm{exp}})^{-1}V_2$, and $(N_1^{\mathrm{exp}}+N_2^{\mathrm{exp}})^{-1}V_{12}$ are the actual physical interaction potentials, whereas for generic increasing values of $N_1,N_2$ \eqref{eq:HN1N2} defines a sequence of auxiliary Hamiltonians acting on larger and larger Hilbert spaces, in which the interaction scales with the particle number with the characteristic mean-field behaviour. Beside, the formal magnitude of the \emph{re-scaled} potential terms match the $O(N_1+N_2)$-magnitude of the kinetic terms, thus making the many-body dynamics governed by $H_{N_1,N_2}$ meaningful at all sizes of the system. The explicit choice of the mean-field pre-factors made in \eqref{eq:HN1N2} can be easily recognised to be the physically appropriate one \cite[Sect.~4]{M-Olg-2016_2mixtureMF}.

 In order for each population to remain relevant in the asymptotic limit, it is assumed that $N_1,N_2\to\infty$ with asymptotically constant and non-zero population ratios $c_1$ and $c_2$, that is,
 \begin{equation}\label{eq:ratios}
  c_k\;:=\;\lim_{N_1,N_2\to\infty}\frac{N_k}{N_1+N_2}\;\in(0,1)\,,\qquad k\in\{1,2\}\,.
 \end{equation}

 It is fairly standard that an ample selection of potentials $U$ and $V_\alpha$ may be made above so as to realise $H_{N_1,N_2}$ on $\cH_{N_1,N_2,\mathrm{sym}}$ self-adjointly and also with lower bound growing at most linearly in the particle number (see \cite[Sect.~2]{M-Olg-2016_2mixtureMF} and \cite{MNO-2017}), a condition that will be implicitly assumed here throughout.

 The Bose mixture initially prepared in a state $\Psi_{N_1,N_2}$, evolves at later times $t>0$ through states of the form
 \begin{equation}\label{eq:SchrEq0}
  \Psi_{N_1,N_2,t}\;:=\;\text{e}^{-\ii t H_{N_1,N_2}}\Psi_{N_1,N_2}\,,
 \end{equation}
 namely the unique solution in the (operator, or form) domain of $H_{N_1,N_2}$ of the Schr\"{o}dinger equation
 \begin{equation}\label{eq:SchrEq}
  \ii\partial_t\Psi_{N_1,N_2,t}\;=\;H_{N_1,N_2}\Psi_{N_1,N_2,t}
 \end{equation}
 with initial datum $\Psi_{N_1,N_2}$. Clearly, for non-zero $V_{12}$ the evolved state $\Psi_{N_1,N_2,t}$ does not factorise into a product separately in the A-variables and B-variables. More importantly, as in physical applications the particle number is very large, ranging from few hundreds to $10^{11}$ and more, the quantum dynamics described by \eqref{eq:SchrEq} has a level of complexity that cannot be matched analytically or numerically.

 Yet, fundamental and informative results can be rigorously obtained on the above time evolution in the asymptotic limit $N_1,N_2\to\infty$ and in the special regime, of major physical interest, of \emph{Bose-Einstein condensation}, where the following has by now repeatedly established with various techniques and refinements \cite{M-Olg-2016_2mixtureMF,Anap-Hott-Hundertmark-2017,DeOliveira-Michelangeli-2016}. If, at $t=0$, and asymptotically as $N_1,N_2\to\infty$,
 \begin{equation}
  \Psi_{N_1,N_2}(x_1,\dots,x_{N_1};y_1,\dots,y_{N_2})\;\approx\;\prod_{i=1}^{N_1}u(x_i)\prod_{r=1}^{N_2}v(y_r)
 \end{equation}
 for some normalised $u,v\in L^2(\mathbb{R}^d)$, where the above approximation is understood in the sense of very small correlation corrections to the purely factorised state, and more precisely in the sense of asymptotically factorised reduced density matrices \cite[Sect.~2-3]{M-Olg-2016_2mixtureMF}, then the same asymptotic factorisation survives at later times, namely,
  \begin{equation}\label{eq:almostfactorised-t}
  \Psi_{N_1,N_2,t}(x_1,\dots,x_{N_1};y_1,\dots,y_{N_2})\;\approx\;\prod_{i=1}^{N_1}u_t(x_i)\prod_{r=1}^{N_2}v_t(y_r)\,.
 \end{equation}
 Moreover, the evolved one-body orbitals $u_t$ and $v_t$ obey the system of coupled non-linear Schr\"{o}dinger equations of Hartree type,
 \begin{equation}\label{eq:Hartree_system}
\begin{split}
\ii\partial_t u_t\;&=\;h_1 u_t + (V_1*|u_t|^2) u_t + c_2 (V_{12}*|v_t|^2) u_t\,, \\
\ii\partial_t v_t\;&=\;h_2 v_t + (V_2*|v_t|^2) v_t + c_1 (V_{12}*|u_t|^2) v_t\,,
\end{split}
\end{equation}
 with initial conditions $u_{t=0}=u$ and $v_{t=0}=v$. Remarkably, monitoring the spatial densities $|u_t|^2$ and $|v_t|^2$ of the cloud of each component in the condensate mixture, the effective evolution equations \eqref{eq:Hartree_system} match precisely the experimental observations.

 With such a picture in mind, the present work focuses on a different, yet intimately connected feature of the mixture's dynamics \eqref{eq:SchrEq0}-\eqref{eq:SchrEq}, namely the dynamical growth of correlations among observables for different particles of the same species and of distinct species. On a state of condensation in which the system is prepared at time zero, the expectation of a product of observables relative to different particles factorises, so that in particular two-point correlation functions vanish; the goal here is to monitor at later times the behaviour of the correlation function and to show that \emph{asymptotically} in $N_1+N_2$ its magnitude remains vanishingly small at any fixed $t$.

 This analysis was already performed in \cite{ES-2008} for single-component condensates, and is re-done here for a multi-component Bose mixture, with an amount of non-trivial adaptations and making certain technical steps simpler and faster. The dynamical formation of correlation in a one-species Bose-Einstein condensate was also rigorously established in \cite{EMS-2008} at the typical short spatial scale of the two-body interaction. At the time the novel approach of \cite{ES-2008} was proposed, that quantitative control on the two-body correlation function also served as the key technical ingredient to rigorously derive, in the one-component case, the effective one-body non-linear dynamics from the linear many-body Schr\"{o}dinger dynamics, within the scheme of the quantum BBGKY hierarchies for the time evolution. As by now the derivation of \eqref{eq:almostfactorised-t}-\eqref{eq:Hartree_system} from \eqref{eq:SchrEq0} is known in great detail for mixtures, and through a variety of even more effective and informative techniques, as already mentioned \cite{M-Olg-2016_2mixtureMF,AO-GPmixture-2016volume,Anap-Hott-Hundertmark-2017,DeOliveira-Michelangeli-2016}, we do not carry on this point of view explicitly, our emphasis being on the two main results formulated below. Still, also in the multi-component case, and within the BBGKY scheme, our analysis leads naturally to the derivation of the effective non-linear dynamics: this line of reasoning has been recently sketched by one of us in \cite{Sant2020MSc}.

 Let us then come to our main findings. We denote by $\langle \cdot \rangle_{N_1, N_2, t}$  the expectation value in the state $\Psi_{N_1, N_2, t}$, i.e., for all bounded operators $\mathcal{O}$ on $\cH_{N_1, N_2}$,
\begin{equation}
    \langle \mathcal{O} \rangle_{N_1, N_2, t} := \langle \Psi_{N_1, N_2, t},\, \mathcal{O}\, \,\Psi_{N_1, N_2, t} \rangle\,,
\end{equation}
writing simply $\langle \mathcal{O} \rangle_{N_1, N_2}$ for the expectation at time zero on the initial state $\Psi_{N_1, N_2}$. Of course, quantum-mechanically relevant operators are the self-adjoint $\mathcal{O}$'s, namely the quantum observables, for which expectations are real numbers. Moreover, for fixed integers $1\leqslant i_1<\cdots<i_{k_1}\leqslant N_1$ (respectively, $1\leqslant j_1<\cdots<j_{k_2}\leqslant N_2$) we denote by $A^{[i_1,\dots,i_{k_1}]}$ (resp., $B^{[j_1,\dots,j_{k_2}]}$) the bounded operator on $\cH_{N_1, N_2}$ acting as a given bounded operator $A$ on $L^2(\mathbb{R}^{k_1 d})$ on the A-type variables $x_{i_1},\dots,x_{i_{k_1}}$ (resp., as a given bounded $B$ on $L^2(\mathbb{R}^{k_2 d})$ on the B-type variables $y_{j_1},\dots,y_{j_{k_2}}$), and trivially, namely tensorised by the identity operator, on all other variables of the same type and all the variables of the other type. Observe that the separate $x$-variables and $y$-variable permutation symmetry of $\Psi_{N_1, N_2, t}$ allows one to re-write
\begin{equation}
 \begin{split}
  & \big\langle A_2^{[i_{n+1},\dots,i_{n+m}]}B_2^{[j_{n+1},\dots,j_{n+m}]}A_1^{[i_{1},\dots,i_{n}]}B_1^{[j_{1},\dots,j_{n}]}\big\rangle_{N_1, N_2, t} \\
  & \qquad =\;\big\langle A_2^{[n+1,\dots,n+m]}B_2^{[n+1,\dots,n+m]}A_1^{[1,\dots,n]}B_1^{[1,\dots,n]}\big\rangle_{N_1, N_2, t}
 \end{split}
\end{equation}
 for any \emph{disjoint} sets $\{ i_{1},\dots,i_{n}\}$ and $\{i_{n+1},\dots,i_{n+m}\}$, as well as any \emph{disjoint} sets $\{ j_{1},\dots,j_{n}\}$ and $\{j_{n+1},\dots,j_{n+m}\}$. Last, we shall use the convention
 \[
  \widehat{\varphi}(p)\,=\,\frac{1}{(2\pi)^{\frac{d}{2}}}\int_{\mathbb{R}^d}\text{e}^{-\ii x\cdot p}\,\varphi(x)\,\ud x
 \]
 for the Fourier transform $\widehat{\varphi}$ of a function $\varphi$; in particular,
 \begin{equation}
  \|\varphi\|_{L^\infty}\,\leqslant\,\|\widehat{\varphi}\|_{L^1}
 \end{equation}
 whenever the r.h.s.~above is finite.
 
 For the correlation functions involving observables on such distinct set of particles of A-type and of B-type we then prove the following.

 \begin{theorem}\label{thm:main1}
  Consider the $N_1+N_2$ binary Bose mixture in $d$ spatial dimensions, with self-adjoint Hamiltonian $H_{N_1,N_2}$ defined in \eqref{eq:HN1N2}, where
  \[
   V_1,V_2\,\in\,L^\infty(\mathbb{R}^d)\,,\quad \widehat{V}_{12}\,\in\,L^1(\mathbb{R}^d)\,.
  \]
   For given $u,v\in L^2(\mathbb{R}^d)$ with $\|u\|_{L^2}=\|v\|_{L^2}=1$, and for $t>0$, let
  \[
   \Psi_{N_1,N_2}\,:=\; u^{\otimes N_1}\otimes v^{\otimes N_2}\,,\quad \Psi_{N_1,N_2,t}\,:=\,\text{e}^{-\ii t H_{N_1,N_2}}\Psi_{N_1,N_2}\,.
  \]
  For fixed positive integers $n,m<\min\{N_1,N_2\}$, let $A_1,B_1$ be two $n$-body bounded operators on $L^2(\mathbb{R}^{nd})$, and let $A_2,B_2$ be two $m$-body bounded operators on $L^2(\mathbb{R}^{md})$.
   Set $N:=N_1+N_2$, $c_k:=N_k/N$ for $k\in\{1,2\}$, and 
%
  \[
   \begin{split}
    c\,&:=\,\min\{c_1,c_2\}\,, \\
    \mathsf{V}\,&:=\, 24\max\{\|V_1\|_{L^\infty},\|V_2\|_{L^\infty},
   \|\widehat{V}_{12}\|_{L^1}\}\,, \\ 
    \alpha_{N,n,m}\,&:=\,\frac{4mn}{9}\,\Big(\frac{8mn}{N}+4(4+3m+3n)\Big)\,.
   \end{split}
  \]
 Then, for any $t\geqslant 0$,
\begin{equation}\label{claimthgrowth2}
\begin{split}
    \big| \big\langle &A_2^{[n+1, ..., n+m]}  B_2^{[n+1, \dots, n+m]}\,\, A_1^{[1, \dots, n]} B_1^{[1, ..., n]} \big\rangle_{N_1, N_2,t} \\ 
    &\quad- \big\langle A_2^{[n+1, ..., n+m]}  B_2^{[n+1, \dots, n+m]} \big\rangle_{N_1, N_2,t} \big\langle  A_1^{[1, \dots, n]}  B_1^{[1, ..., n]}\big\rangle_{N_1, N_2, t}\big|\\ 
    & \leqslant \frac{\:\alpha_{N,n,m}\,}{\,c^2 N} \,\|A_1\|_{\mathrm{op}}\|A_2\|_{\mathrm{op}}\|B_1\|_{\mathrm{op}}\|B_2\|_{\mathrm{op}}\,(\emph{e}^{\mathsf{V}t}-1) \,.
\end{split}
\end{equation}
 \end{theorem}

 The bound \eqref{claimthgrowth2} shows indeed that at every fixed time the correlation functions built upon the considered observables acting on disjoint sets of particles remain vanishingly small asymptotically in the total number of particles, with a quantitative $O(N^{-1})$ vanishing rate. In particular, this control is exact at $t=0$, as it gives indeed an identically zero correlation function, consistently with the fact that the initial state $\Psi_{N_1,N_2}$ was chosen for concreteness to be exactly factorised (uncorrelated).

 The exponential deterioration in time of the r.h.s.~of \eqref{claimthgrowth2} is clearly an artificial by-product of the (Gr\o{}nwall-like) estimates performed in the proof: obviously, the l.h.s.~of \eqref{claimthgrowth2} is trivially bounded by twice the product of the operator norms, uniformly in $t$. In fact, Theorem \ref{thm:main1} guarantees the asymptotic vanishing in $N$ of the considered correlation functions for the Bose mixture at least up to times
 \[
  t\,\leqslant\,(1-\varepsilon)\,\mathsf{V}^{-1}\log N
 \]
 for arbitrarily small, positive (but non-zero) $\varepsilon$.

 It also goes without saying that we could have stated Theorem \ref{thm:main1} replacing the correlation estimate \eqref{claimthgrowth2} with a completely analogous one where the operators $A_1$ and $A_2$ act on disjoint sets of $n_1$ and $m_1$ A-particles, and the operators $B_1$ and $B_2$ act on disjoint sets of $m_1$ and $m_2$ B-particles. The same $(c^2N)^{-1}$-vanishing in the total particle number would emerge, with a more clumsy expression in $n_1,n_2,m_1,m_2$ replacing the numerical pre-factor $\alpha_{N,n,m}$.

 Theorem \ref{thm:main1} is noticeable both for the control of correlations it provides, and for the specific technique used for its proof. It is based on a version for multi-component Bose mixtures of Lieb-Robinson-type bounds originally found in \cite{ES-2008} for single-component condensates. In their original version, Lieb-Robinson bounds are theoretical upper limits on the finite propagation speed of certain information in non-relativistic quantum systems, concretely speaking spin lattices \cite[Sect.~6.2.1]{Bratteli-Robinson-2}. They have had a wide spectrum of applications and consequences for the dynamics of many-body quantum system \cite{Nachtergaele-Sims-2010-LR,Sims-2011-LR}.

 In the present analysis for multi-component Bose mixtures we establish a control of Lieb-Robinson type on the size of the commutator between operators of the form
 \[
  A_1^{[i_1,\dots,i_{n_1}]}B_1^{[j_1,\dots,j_{n_2}]}\quad\textrm{and}\quad A_2^{[i_{n_1+1},\dots,i_{n_1+m_1}]}B_2^{[j_{n_2+1},\dots,j_{n_2+m_2}]}
 \]
for \emph{disjoint} variable sets $\{i_1,\dots,i_{n_1}\}$ and $\{i_{n_1+1},\dots,i_{n_1+m_1}\}$ and \emph{disjoint} variable sets $\{j_1,\dots,j_{n_2}\}$ and $\{j_{n_2+1},\dots,j_{n_2+m_2}\}$. Obviously, such two operators commute because they act on different variables, but when one of the two is evolved according to the dynamics governed by $H_{N_1,N_2}$ the commutativity is destroyed and the norm of the new commutator then measures the rate at which the disturbances caused by the interaction propagate. The goal is then to show that such an effect, which takes place with a finite speed of propagation, is vanishingly small asymptotically in the total number of particles. This is encoded in our second main result, which has abstract relevance per se and, as said, is instrumental for the proof of Theorem \ref{thm:main1}. Once again, permutation symmetry allows one to re-label the variables in increasing order.
 
 \begin{theorem}\label{thm:LR}
    Consider the $N_1+N_2$ binary Bose mixture in $d$ spatial dimensions, with self-adjoint Hamiltonian $H_{N_1,N_2}$ defined in \eqref{eq:HN1N2}, where 
  \[
   V_1,V_2\,\in\,L^\infty(\mathbb{R}^d)\,,\quad \widehat{V}_{12}\,\in\,L^1(\mathbb{R}^d)\,.
  \]
  Set 
  \[
   \mathcal{V}\,:=12\,\max\{\|V_1\|_{L^\infty},\|V_2\|_{L^\infty},
   \|\widehat{V}_{12}\|_{L^1}\}\,.
  \]
  For fixed positive integers $n_1,m_1< N_1$ and $n_2,m_2< N_2$, consider the bounded operators 
  $A_1,A_2,B_1,B_2$ respectively on $L^2(\mathbb{R}^{n_1d}),L^2(\mathbb{R}^{m_1d}),L^2(\mathbb{R}^{n_2d}),L^2(\mathbb{R}^{m_2d})$\,.
  Then, for any $t\geqslant 0$,
  \begin{equation}\label{eq:LRbound-general}
   \begin{split}
    & \big\| \big[ \,A_2^{[{n_1+1},\dots,{n_1+m_1}]}B_2^{[{n_2+1},\dots,{n_2+m_2}]}\,,\,\text{e}^{\ii t H_{N_1,N_2}} A_1^{[1,\dots,{n_1}]} B_1^{[1,\dots,{n_2}]} \text{e}^{-\ii t H_{N_1,N_2}}\, \big] \big\|_{\mathrm{op}} \\
    &\qquad\leqslant\;
    \|A_1\|_{\mathrm{op}}\|A_2\|_{\mathrm{op}}\|B_1\|_{\mathrm{op}}\|B_2\|_{\mathrm{op}}\,\frac{\,(m_1+m_2)(n_1+n_2)\,}{3\,c\,N}(\text{e}^{\mathcal{V} t}-1)\,.
   \end{split}
  \end{equation}
 \end{theorem}

 The proof of the Lieb-Robinson-like bound \eqref{eq:LRbound-general} is presented in Section \ref{sec:LR}. Based on it, Theorem \ref{thm:main1} is then proved in Section \ref{sec:corr}.

 Both Theorems \ref{thm:main1} and \ref{thm:LR} are somewhat rigid in two key assumptions, as will be clear from their proofs: the boundedness of the interaction potentials (plus the integrability of the Fourier transform of the inter-species interaction), and the mean-field scaling in the many-body Hamiltonian $H_{N_1,N_2}$ as defined in \eqref{eq:HN1N2}. This does not spoil the physical meaningfulness of such rigorous results, though. Indeed, typical Coulomb or van der Waals two-body potentials are not covered, but one should consider that eventually their local divergence is corrected by real-world finite-size behaviours; beside, it is conceivable that singular potentials may be included in the present context at the price of some kind of higher-regularity control of the quantities of interest, in the same spirit, quite natural in the related literature, of \cite{EESY2006,ESYinvent,MS-2012-BosonStar,Anap-Hott-Hundertmark-2017}.

 As for the possibility of implementing scalings in $N$ that are surely physically more realistic, like replacing the mean-field scaling $N^{-1}V(x)$ with the Gross-Pitaevskii scaling $N^2 V(Nx)$ in $d=3$ dimensions \cite{am_GPlim}, it should be emphasised, as recently pointed out in detail in an analogous model in \cite{MO-2018-spin-spin}, that when one restores the physical units in mean-field bounds of the type emerged above and evaluates them plugging in the actual values of the experimental parameters, the fidelity of the mean-field description for the whole typical duration of the time evolution is already impressively good.

 We do not carry on the analysis of such generalisations here, sticking to the novelty of establishing and exploiting Lieb-Robinson-like bounds for the control of correlations in the dynamics of a (mean-field) Bose mixture.

 \section{Lieb-Robinson bounds}\label{sec:LR}

 The general idea for the proof of Theorem \ref{thm:LR} is to manipulate the l.h.s.~of \eqref{eq:LRbound-general}, suitably re-writing it as integral of its time derivative, in such a way to obtain the structure of an integral Gr\o{}nwall-like estimate. This means that the size of the l.h.s.~of \eqref{eq:LRbound-general} is to be controlled by terms that essentially reproduce the integral in time of the same l.h.s., plus terms that grow from zero linearly in time. It is crucial that the latter are shown to be $O(N^{-1})$-small in the particle number, in order to get the asymptotic $N^{-1}$-smallness of the final result. To this aim, it is convenient to introduce a modified dynamics that effectively decouples the fixed number of particles under consideration from the cloud of the rest of the mixture.

 For convenience we write 
 \begin{equation}
  \mathcal{B}'_n\;:=\;\mathcal{B}(L^2(\mathbb{R}^{nd}))\setminus\{\mathbb{O}\}
 \end{equation}
 for the non-zero elements of the $C^*$-algebra of bounded operators acting on the Hilbert space $L^2(\mathbb{R}^{nd})$ of $n$ particles in $d$ dimensions. Moreover, for positive integers $n_1,m_1< N_1$ and $n_2,m_2< N_2$, and for given $A_1\in\mathcal{B}'_{n_1}$, $A_2\in\mathcal{B}'_{m_1}$, $B_2\in\mathcal{B}'_{n_2}$, $B_2\in\mathcal{B}'_{m_2}$, we introduce the short-hand
 \begin{equation}\label{shorthand2}
  \begin{array}{lcrclcr}
   A^{\mathbf{1}} \!\!\!& \equiv & \!\!\!A_1^{[1, \dots, n_1]}\,, & \qquad & A^{\mathbf{2}} \!\!\!& \equiv & \!\!\!A_2^{[n_1+1, \dots, n_1+m_1]}\,, \\
   B^{\mathbf{1}} \!\!\!& \equiv & \!\!\!B_1^{[1, \dots, n_2]}\,, & \qquad & B^{\mathbf{2}} \!\!\!& \equiv & \!\!\!A_2^{[n_2+1, \dots, n_2+m_2]}\,,
  \end{array}
 \end{equation}
 for such operators on $\mathcal{H}_{N_1,N_2}$ denoted according to the convention already presented in Section \ref{intro}. Next, we define, for $t\geqslant 0$,
 \begin{equation}\label{eq:Fquantity}
  \mathcal{F}_{\substack{ m_1,m_2 \\ n_1,n_2} }(t)\;:=\;\sup_{ \substack{  A_1\in\mathcal{B}'_{n_1}, \,A_2\in\mathcal{B}'_{m_1} \\  B_1\in\mathcal{B}'_{n_2}, \,B_2\in\mathcal{B}'_{m_2}}} \frac{\;\big\| \big[ A^{\mathbf{2}}B^{\mathbf{2}} , \text{e}^{\ii t H_{N_1,N_2}}A^{\mathbf{1}}B^{\mathbf{1}}\text{e}^{-\ii t H_{N_1,N_2}}\big] \big\|_{\mathrm{op}}\,}{\|A^{\mathbf{1}}\|_{\mathrm{op}}\|A^{\mathbf{2}}\|_{\mathrm{op}}\|B^{\mathbf{1}}\|_{\mathrm{op}}\|B^{\mathbf{2}}\|_{\mathrm{op}}}\,.
 \end{equation}
  The desired Lieb-Robinson bound \eqref{eq:LRbound-general} then follows if one proves that
  \begin{equation}\label{eq:whatwewant}
   \mathcal{F}_{\substack{ m_1,m_2 \\ n_1,n_2} }(t)\;\leqslant\;\frac{\,(m_1+m_2)(n_1+n_2)\,}{3\,c\,N}(\text{e}^{\mathcal{V} t}-1)
  \end{equation}
  at any $t\geqslant 0$.

  On the other hand, we introduce the modified Hamiltonian
 \begin{equation}\label{modH2}
    \begin{split}
        H_{N_1, N_2}^{(n_1, n_2)}\;&:=\; H_{N_1, N_2} \\
        &\qquad -\frac{1}{N_1} \sum_{i=1}^{n_1}\sum_{j=n_1+1}^{N_1} V_1(x_i-x_j) -\frac{1}{N_2}\sum_{r=1}^{n_2}\sum_{p=n_2+1}^{N_2} V_2(y_r-y_p) \\
        & \qquad -\frac{1}{N_1+N_2} \sum_{i=1}^{n_1}\sum_{p=n_2+1}^{N_2} V_{12}(x_i-y_p) \\
        &\qquad\qquad -\frac{1}{N_1+N_2} \sum_{j=n_1+1}^{N_1}\sum_{r=1}^{n_2} V_{12}(x_j-y_r)
    \end{split}
\end{equation}
 modelling the dynamics where the sole interactions left are those involving the first $n_1$ A-particles and the first $n_2$ B-particles. By construction, both $A^{\mathbf{1}}B^{\mathbf{1}}$ and the time-dependent operator
 \[
   \text{e}^{-\ii t H_{N_1,N_2}^{(n_1, n_2)}}A^{\mathbf{1}}B^{\mathbf{1}}\text{e}^{\ii t H_{N_1,N_2}^{(n_1, n_2)}}
 \]
 act non-trivially only on the two sets of variables $x_1,\dots,x_{n_1}$ and $y_1,\dots,y_{n_2}$, and actually with the same operator norm. Thus, the supremum in \eqref{eq:Fquantity} can be equivalently chosen to run on the latter type of operators, and one can re-write
 \begin{equation}
  \mathcal{F}_{\substack{ m_1,m_2 \\ n_1,n_2} }(t)\;=\;\sup_{ \substack{  A_1\in\mathcal{B}'_{n_1}, \,A_2\in\mathcal{B}'_{m_1} \\  B_2\in\mathcal{B}'_{n_2}, \,B_2\in\mathcal{B}'_{m_2}}}\frac{\big\|\mathcal{G}(t)\big\|_{\mathrm{op}}}{\|A^{\mathbf{1}}\|_{\mathrm{op}}\|A^{\mathbf{2}}\|_{\mathrm{op}}\|B^{\mathbf{1}}\|_{\mathrm{op}}\|B^{\mathbf{2}}\|_{\mathrm{op}}}
 \end{equation}
 with
 \begin{equation}
  \mathcal{G}(t)\;:=\;\big[ A^{\mathbf{2}}B^{\mathbf{2}} , \text{e}^{\ii t H_{N_1,N_2}}\text{e}^{-\ii t H_{N_1,N_2}^{(n_1, n_2)}}A^{\mathbf{1}}B^{\mathbf{1}}\text{e}^{\ii t H_{N_1,N_2}^{(n_1, n_2)}}\text{e}^{-\ii t H_{N_1,N_2}}\big]\,.
 \end{equation}

  Now, setting
  \begin{equation}
    \mathcal{H}^{(n_1, n_2)}(t) \;:=\; \text{e}^{\ii t H_{N_1, N_2}} (H_{N_1, N_2}- H_{N_1, N_2}^{(n_1, n_2)}) \text{e}^{-\ii t H_{N_1, N_2}}\,,
\end{equation}
 and with the short-hand
 \begin{equation}
    (A^{\mathbf{1}}B^{\mathbf{1}})_t \;\equiv\; \text{e}^{\ii t H_{N_1,N_2}}\text{e}^{-\ii t H_{N_1,N_2}^{(n_1, n_2)}}A^{\mathbf{1}}B^{\mathbf{1}}\text{e}^{\ii t H_{N_1,N_2}^{(n_1, n_2)}}\text{e}^{-\ii t H_{N_1,N_2}}\,,
\end{equation}
  we find
\begin{equation}\label{dgbin}
   \frac{\ud \mathcal{G}(t)}{\ud t} \;=\; [A_2 B_2, [i \mathcal{H}^{(n_1, n_2)}(t),(A^{\mathbf{1}}B^{\mathbf{1}})_t]]\,.
\end{equation}
  Moreover, the Jacobi identity in the r.h.s.~of \eqref{dgbin} yields
  \begin{equation}\label{eq:dGtJac}
   \frac{\ud \mathcal{G}(t)}{\ud t}\;=\;\ii \big[ \mathcal{H}^{(n_1, n_2)}(t), \mathcal{G}(t)\big] - \ii\big[(A^{\mathbf{1}}B^{\mathbf{1}})_t, \big[ A^{\mathbf{2}}B^{\mathbf{2}}, \mathcal{H}^{(n_1, n_2)}(t) \big] \big]\,.
  \end{equation}

  It follows by standard arguments (see, e.g., \cite[Sect.~4.2.2]{Sant2020MSc}) that $\{\mathcal{H}^{(n_1, n_2)}(t),t\in\mathbb{R}\}$ is a family of self-adjoint operators on $\mathcal{H}_{N_1,N_2}$ generating a two-parameter group $\{\mathcal{U}^{(n_1, n_2)}(t, s)|s,t\in\mathbb{R}\}$ of unitary transformations satisfying, for each $s,t$,
  \begin{equation}\label{groupstru2}
    \begin{cases}
    \ii \partial_t\, \mathcal{U}^{(n_1, n_2)}(t,s) \,=\, \mathcal{H}^{(n_1, n_2)}(t)\, \mathcal{U}^{(n_1, n_2)}(t, s)\,, \\
    \mathcal{U}^{(n_1, n_2)}(s,s) \,=\, \mathbbm{1}\,.
    \end{cases}
\end{equation}
  From the group structure $\mathcal{U}^{(n_1, n_2)}(r,s)\mathcal{U}^{(n_1, n_2)}(s,t)=\mathcal{U}^{(n_1, n_2)}(r,t)$ and from the identity $\mathcal{U}^{(n_1, n_2)}(t,t)=\mathbbm{1}$, valid $\forall$ $r,s,t\in\mathbb{R}$,  one finds $ \mathcal{U}^{(n_1, n_2)}(s,t)^*=\mathcal{U}^{(n_1, n_2)}(t,s)$. Thus, from \eqref{groupstru2}, and taking the adjoint, one deduces, for all $t$,
  \begin{equation}\label{eq:dUn1n2}
   \begin{split}
    \frac{\ud}{\ud t}\, \mathcal{U}^{(n_1, n_2)}(t,0) \;&=\; -\ii\,\mathcal{H}^{(n_1, n_2)}(t)\, \mathcal{U}^{(n_1, n_2)}(t, 0)\,, \\
    \frac{\ud}{\ud t}\, \mathcal{U}^{(n_1, n_2)}(0,t) \;&=\; \ii \,\mathcal{U}^{(n_1, n_2)}(0, t)\mathcal{H}^{(n_1, n_2)}(t)\,.
   \end{split}
  \end{equation}

  By means of \eqref{eq:dGtJac} and \eqref{eq:dUn1n2} it is now possible to perform a more convenient differentiation in time, namely with $\mathcal{G}(t)$ \emph{dressed} with the unitaries $\mathcal{U}^{(n_1, n_2)}(t,0)$. This yields
  \begin{equation}
   \begin{split}
     \frac{\ud}{\ud t}\,&\mathcal{U}^{(n_1, n_2)}(t,0)\,\mathcal{G}(t)\,\mathcal{U}^{(n_1, n_2)}(0,t) \\
     &=\;\mathcal{U}^{(n_1, n_2)}(t, 0)\, \big[(A^{\mathbf{1}}B^{\mathbf{1}})_t, \big[A^{\mathbf{2}}B^{\mathbf{2}}, \mathcal{H}^{(n_1, n_2)}(t) \big]\big]\,\mathcal{U}^{(n_1, n_2)}(0, t)\,.
   \end{split}
  \end{equation}
  Integration in time over $[0,t]$ clearly gives
  \[
   \begin{split}
    &\mathcal{U}^{(n_1, n_2)}(t,0)\,\mathcal{G}(t)\,\mathcal{U}^{(n_1, n_2)}(0,t) \\
    &\qquad =\;\int_0^t\ud s\,\mathcal{U}^{(n_1, n_2)}(s, 0)\, \big[(A^{\mathbf{1}}B^{\mathbf{1}})_s, \big[A^{\mathbf{2}}B^{\mathbf{2}}, \mathcal{H}^{(n_1, n_2)}(s) \big]\big]\,\mathcal{U}^{(n_1, n_2)}(0, s)\,,
   \end{split}
  \]
 whence, taking the operator norm on both sides and exploiting unitarity,
  \begin{equation}\label{gnorm2}
\begin{split}
     &\|\mathcal{G}(t)\|_{\mathrm{op}} \;\leqslant \; 2 \big\|(A^{\mathbf{1}}B^{\mathbf{1}})_t\big\|_{\mathrm{op}} \int_0^t  \ud s\,\big\|[A^{\mathbf{2}}B^{\mathbf{2}}, \mathcal{H}^{(n_1, n_2)}(s)]\big\| _{\mathrm{op}}
     \\ 
     &=\; 2 \big\|A^{\mathbf{1}}B^{\mathbf{1}}\|_{\mathrm{op}} \int_0^t  \ud s\,\big\|\big[A^{\mathbf{2}}B^{\mathbf{2}}, \text{e}^{\ii s H_{N_1, N_2}} (H_{N_1, N_2}- H_{N_1, N_2}^{(n_1, n_2)}) \text{e}^{-\ii s H_{N_1, N_2}}\big]\big\| _{\mathrm{op}}\,.
\end{split}
\end{equation}

  The difference $H_{N_1, N_2}- H_{N_1, N_2}^{(n_1, n_2)}$ is crucial in \eqref{gnorm2}: it is going to be the source of $O(N^{-1})$-small contributions in the Gr\o{}nwall-like estimate we are building up. It contains all interaction terms between the considered $n_1$ A-particles and $n_2$ B-particles on the one hand, and all the particles of the surrounding cloud on the other. It is convenient to split their sum as
  \begin{equation}\label{modHabcd}
\begin{split}
  & H_{N_1, N_2} -  H_{N_1, N_2}^{(n_1, n_2)} \;=\;   \mathcal{A}_{1,\, n_1+1}^{n_1,\, n_1+m_1} +  \mathcal{A}_{1, \,n_1+m_1+1}^{n_1,\,N_1 } +   \mathcal{B}_{1,\, n_2+1}^{n_2,\, n_2+m_2}  \\ 
  &\quad + \mathcal{B}_{1,\,n_2+m_2+1}^{n_1, \,N_2} + \mathcal{C}_{1,\,n_2+1}^{n_1,\,n_2+m_2} + \mathcal{C}_{1,\,n_2+m_2+1}^{n_1,\,N_2}
    +  \mathcal{D}_{n_1+1,\, 1}^{n_1+m_1,\,n_2} + \mathcal{D}_{n_1+m_1+1, \,1}^{N_1,\,n_2}\,,
\end{split}
\end{equation}
  where
  \begin{eqnarray}
   \mathcal{A}_{1,\, n_1+1}^{n_1,\, n_1+m_1} \!\!&:=&\!\! \frac{1}{N_1}\sum_{i=1}^{n_1} \sum_{j=n_1+1}^{n_1+m_1} V_1(x_i-x_j)\,, \label{A1-outof8}\\
   \mathcal{A}_{1, \,n_1+m_1+1}^{n_1,\,N_1 } \!\!&:=&\!\!\frac{1}{N_1}\sum_{i=1}^{n_1} \sum_{j=n_1+m_1+1}^{N_1} V_1(x_i-x_j)\,, \\
   \mathcal{B}_{1,\, n_2+1}^{n_2,\, n_2+m_2} \!\!&:=&\!\!\frac{1}{N_2}\sum_{r=1}^{n_2} \sum_{p=n_2+1}^{n_2+m_2} V_2(y_r-y_p)\,, \\
   \mathcal{B}_{1,\,n_2+m_2+1}^{n_1, \,N_2}\!\!&:=&\!\!\frac{1}{N_2}\sum_{r=1}^{n_2} \sum_{p=n_2+m_2+1}^{N_2}  V_2(y_r-y_p)\,, \\
   \mathcal{C}_{1,\,n_2+1}^{n_1,\,n_2+m_2}\!\!&:=&\!\! \frac{1}{N_1+N_2} \sum_{i=1}^{n_1} \sum_{p=n_2+1}^{n_2+m_2} V_{12}(x_i-y_p)\,, \\
   \mathcal{C}_{1,\,n_2+m_2+1}^{n_1,\,N_2}\!\!&:=&\!\! \frac{1}{N_1+N_2} \sum_{i=1}^{n_1} \sum_{p=n_2+m_2+1}^{N_2} V_{12}(x_i-y_p)\,, \\
   \mathcal{D}_{n_1+1,\, 1}^{n_1+m_1,\,n_2}\!\!&:=&\!\!\frac{1}{N_1+N_2} \sum_{j=n_1+1}^{n_1+m_1} \sum_{r=1}^{n_2} V_{12}(x_j-y_r)\,, \\
   \mathcal{D}_{n_1+m_1+1, \,1}^{N_1,\,n_2}\!\!&:=&\!\!\frac{1}{N_1+N_2} \sum_{j=n_1+m_1+1}^{N_1} \sum_{r=1}^{n_2} V_{12}(x_j-y_r)\,. \label{D2-outof8}
  \end{eqnarray}
  Thus, plugging \eqref{modHabcd} into \eqref{gnorm2}, we obtain
  \begin{equation}\label{tosplit}
   \frac{\;\|\mathcal{G}(t)\|_{\mathrm{op}}}{\;\big\|A^{\mathbf{1}}\|_{\mathrm{op}}\big\|B^{\mathbf{1}}\|_{\mathrm{op}}}\;\leqslant\;2\sum_{\mathcal{J}}\int_0^t  \ud s\,\big\|\big[A^{\mathbf{2}}B^{\mathbf{2}}, \text{e}^{\ii s H_{N_1, N_2}} \mathcal{J} \text{e}^{-\ii s H_{N_1, N_2}}\big]\big\| _{\mathrm{op}}\,,
  \end{equation}
 where $\mathcal{J}$ runs over the eight operators \eqref{A1-outof8}-\eqref{D2-outof8}.

  The various terms in the r.h.s.~of \eqref{tosplit} are controlled as follows. Four of them, those containing a number of interaction pairs which does \emph{not} scale with $N_1$ and $N_2$, are estimated trivially, like
  \begin{equation}\label{oddterms1}
    \begin{split}
        &2\int_0^t  \ud s\,\big\|\big[A^{\mathbf{2}}B^{\mathbf{2}}, \text{e}^{\ii s H_{N_1, N_2}} \mathcal{A}_{1,\, n_1+1}^{n_1,\, n_1+m_1} \text{e}^{-\ii s H_{N_1, N_2}}\big]\big\| _{\mathrm{op}} \\
        &\qquad \leqslant\;   \frac{4\big\|A^{\mathbf{2}}\|_{\mathrm{op}}\big\|B^{\mathbf{2}}\|_{\mathrm{op}}}{N_1}\sum_{i=1}^{n_1} \sum_{j=n_1+1}^{n_1+m_1} \|V_1\|_{L^\infty}\int_0^t \ud s  \\
        &\qquad =\;  \frac{4 n_1 m_1}{N_1} \big\|A^{\mathbf{2}}\|_{\mathrm{op}}\big\|B^{\mathbf{2}}\|_{\mathrm{op}} \|V_1\|_{L^\infty} t\,.
     \end{split}
\end{equation}  
  Analogously, the contributions from the terms with $\mathcal{B}_{1,\, n_2+1}^{n_2,\, n_2+m_2}$, $\mathcal{C}_{1,\,n_2+1}^{n_1,\,n_2+m_2}$, and $\mathcal{D}_{n_1+1,\, 1}^{n_1+m_1,\,n_2}$ are bounded by $\big\|A^{\mathbf{2}}\|_{\mathrm{op}}\big\|B^{\mathbf{2}}\|_{\mathrm{op}}$ times, respectively,
  \begin{equation}\label{oddterms1again}
   \frac{4 n_2 m_2}{N_1} \|V_2\|_{L^\infty} t\,,\quad \frac{4 n_1 m_2}{N_1+N_2} \|V_{12}\|_{L^\infty} t\,,\quad\frac{4 n_2 m_1}{N_1+N_2} \|V_{12}\|_{L^\infty} t\,.
  \end{equation}

  In the remaining terms the $O(N^{-1})$ mean-field pre-factors are compensated by the large $O(N)$ number of interaction pairs, and it is therefore not possible to pull out the above vanishing behaviour in the particle number. Such terms are rather controlled in terms of the original $ \mathcal{F}_{\substack{ m_1,m_2 \\ n_1,n_2} }(t)$. Thus,
\begin{equation}\label{eq:contr4}
\begin{split}
    &2\int_0^t  \ud s\,\big\|\big[A^{\mathbf{2}}B^{\mathbf{2}}, \text{e}^{\ii s H_{N_1, N_2}} \mathcal{A}_{1, \,n_1+m_1+1}^{n_1,\,N_1 } \text{e}^{-\ii s H_{N_1, N_2}}\big]\big\| _{\mathrm{op}} \\
    & \leqslant\; \frac{2}{N_1} \sum_{i=1}^{n_1} \sum_{j=n_1+m_1+1}^{N_1} \int_0^t \ud s\, \big\|\big[A^{\mathbf{2}}B^{\mathbf{2}}, \text{e}^{\ii s H_{N_1, N_2}}V_1(x_i-x_j) \text{e}^{-\ii s H_{N_1, N_2}}\big]\big\| _{\mathrm{op}} \\
    &= \;2\; \frac{n_1(N_1-n_1-m_1)}{N_1} \times \\
    &\qquad \times \int_0^t \ud s\, \big\|\big[A^{\mathbf{2}}B^{\mathbf{2}}, \text{e}^{\ii s H_{N_1, N_2}}V_1(x_1-x_{n_1+m_1+1}) \text{e}^{-\ii s H_{N_1, N_2}}\big]\big\| _{\mathrm{op}} \\
    &\leqslant\;2 n_1 \big\|A^{\mathbf{2}}\|_{\mathrm{op}}\big\|B^{\mathbf{2}}\|_{\mathrm{op}}\|V_1\|_{L^\infty}\int_0^t \ud s\,\mathcal{F}_{\substack{ m_1,m_2 \\ 2,0} }(s)\,,
\end{split}
\end{equation}
 where in the third line above we used the permutation symmetry for $x$-type variables, and in the last step we used $\frac{N_1-n_1-m_1}{N_1}\leqslant 1$ and the identity (see definition \eqref{eq:Fquantity} at the beginning)
 \begin{equation}
 \mathcal{F}_{\substack{ m_1,m_2 \\ 2,0} }(s)\;=\;\sup_{ \substack{  A_1\in\mathcal{B}'_{2}, \,A_2\in\mathcal{B}'_{m_1} \\  B_2\in\mathcal{B}'_{m_2}}} \frac{\;\big\| \big[ A^{\mathbf{2}}B^{\mathbf{2}} , \text{e}^{\ii s H_{N_1,N_2}} A^{\mathbf{1}} \text{e}^{-\ii s H_{N_1,N_2}}\big] \big\|_{\mathrm{op}}\,}{\|A^{\mathbf{2}}\|_{\mathrm{op}}\|B^{\mathbf{2}}\|_{\mathrm{op}}\|A^{\mathbf{1}}\|_{\mathrm{op}}}\,.
 \end{equation}
  Analogously,
  \begin{equation}\label{eq:boundagain1}      
   \begin{split}
    &2\int_0^t  \ud s\,\big\|\big[A^{\mathbf{2}}B^{\mathbf{2}}, \text{e}^{\ii s H_{N_1, N_2}} \mathcal{B}_{1,\,n_2+m_2+1}^{n_1, \,N_2} \text{e}^{-\ii s H_{N_1, N_2}}\big]\big\| _{\mathrm{op}} \\
    &\leqslant\;2 n_2 \big\|A^{\mathbf{2}}\|_{\mathrm{op}}\big\|B^{\mathbf{2}}\|_{\mathrm{op}}\|V_2\|_{L^\infty}\int_0^t \ud s\,\mathcal{F}_{\substack{ m_1,m_2 \\ 0,2} }(s)\,.
   \end{split}
  \end{equation}

  The remaining contributions, from the terms with $\mathcal{C}_{1,\,n_2+m_2+1}^{n_1,\,N_2}$ and $\mathcal{D}_{n_1+m_1+1, \,1}^{N_1,\,n_2}$, require a more elaborate strategy as compared to \eqref{eq:contr4}, for the multiplicative operators $V_{12}(x_i-y_p)$ cannot be interpreted, unlike $V_1(x_1-x_{n_1+m_1+1})$ in \eqref{eq:contr4}, as a one-body observable acting on an A-particle or a B-particle only. Instead, we write
  \begin{equation}\label{eq:withFtransf}
   \begin{split}
    & \big\|\big[A^{\mathbf{2}}B^{\mathbf{2}}, \text{e}^{\ii s H_{N_1, N_2}} V_{12}(x_i-y_p)\text{e}^{-\ii s H_{N_1, N_2}}\big]\big\| _{\mathrm{op}} \\
    & \qquad\leqslant \int_{\mathbb{R}^d}\ud q\, |\widehat{V}_{12}(q)|\,\big\|\big[A^{\mathbf{2}}B^{\mathbf{2}}, \text{e}^{\ii s H_{N_1, N_2}}\text{e}^{\ii q x_i}\otimes \text{e}^{-\ii q y_p}\text{e}^{-\ii s H_{N_1, N_2}}\big]\big\| _{\mathrm{op}} \\
    & \qquad\leqslant\;\big\|A^{\mathbf{2}}\|_{\mathrm{op}}\big\|B^{\mathbf{2}}\|_{\mathrm{op}}\|\widehat{V}_{12}\|_{L^1}\,\mathcal{F}_{\substack{ m_1,m_2 \\ 1,1} }(s)\,,
   \end{split}
  \end{equation}
  having used, from \eqref{eq:Fquantity},
   \begin{equation}
 \mathcal{F}_{\substack{ m_1,m_2 \\ 1,1} }(s)\;=\;\sup_{ \substack{  A_1\in\mathcal{B}'_{1}, \,A_2\in\mathcal{B}'_{m_1} \\   B_1\in\mathcal{B}'_{1},\, B_2\in\mathcal{B}'_{m_2}}} \frac{\;\big\| \big[ A^{\mathbf{2}}B^{\mathbf{2}} , \text{e}^{\ii s H_{N_1,N_2}} A^{\mathbf{1}} B^{\mathbf{1}}\text{e}^{-\ii s H_{N_1,N_2}}\big] \big\|_{\mathrm{op}}\,}{\|A^{\mathbf{1}}\|_{\mathrm{op}}\|A^{\mathbf{2}}\|_{\mathrm{op}}\|B^{\mathbf{1}}\|_{\mathrm{op}}\|B^{\mathbf{2}}\|_{\mathrm{op}}}\,.
 \end{equation}
  Then \eqref{eq:withFtransf} implies
 \begin{equation}\label{eq:boundfinelista1}
  \begin{split}
   & 2\int_0^t  \ud s\,\big\|\big[A^{\mathbf{2}}B^{\mathbf{2}}, \text{e}^{\ii s H_{N_1, N_2}} \mathcal{C}_{1,\,n_2+m_2+1}^{n_1,\,N_2} \text{e}^{-\ii s H_{N_1, N_2}}\big]\big\| _{\mathrm{op}} \\
   & \qquad\leqslant\;\frac{2}{N_1+N_2}\sum_{i=1}^{n_1} \sum_{p=n_2+m_2+1}^{N_2} \times \\
   &\qquad\qquad\times\int_0^t \ud s\,\big\|\big[A^{\mathbf{2}}B^{\mathbf{2}}, \text{e}^{\ii s H_{N_1, N_2}} V_{12}(x_i-y_p)\text{e}^{-\ii s H_{N_1, N_2}}\big]\big\| _{\mathrm{op}} \\
   & \qquad\leqslant\;\frac{ 2 n_1 (N_2-n_2-m_2)}{N_1+N_2}\big\|A^{\mathbf{2}}\|_{\mathrm{op}}\big\|B^{\mathbf{2}}\|_{\mathrm{op}}\|\widehat{V}_{12}\|_{L^1}\int_0^t \ud s\,\mathcal{F}_{\substack{ m_1,m_2 \\ 1,1} }(s) \\
   & \qquad\leqslant\;2 n_1(1-c)\big\|A^{\mathbf{2}}\|_{\mathrm{op}}\big\|B^{\mathbf{2}}\|_{\mathrm{op}}\|\widehat{V}_{12}\|_{L^1}\int_0^t \ud s\,\mathcal{F}_{\substack{ m_1,m_2 \\ 1,1} }(s)\,,
  \end{split}
 \end{equation}
  In an analogous fashion,
  \begin{equation}\label{eq:boundfinelista2}
   \begin{split}
     & 2\int_0^t  \ud s\,\big\|\big[A^{\mathbf{2}}B^{\mathbf{2}}, \text{e}^{\ii s H_{N_1, N_2}} \mathcal{D}_{n_1+m_1+1, \,1}^{N_1,\,n_2} \text{e}^{-\ii s H_{N_1, N_2}}\big]\big\| _{\mathrm{op}} \\
      & \qquad \leqslant\;2 n_2(1-c)\big\|A^{\mathbf{2}}\|_{\mathrm{op}}\big\|B^{\mathbf{2}}\|_{\mathrm{op}}\|\widehat{V}_{12}\|_{L^1}\int_0^t \ud s\,\mathcal{F}_{\substack{ m_1,m_2 \\ 1,1} }(s)\,.
   \end{split}
  \end{equation}

  Plugging the bounds  \eqref{oddterms1}-\eqref{eq:contr4}, \eqref{eq:boundagain1}, and \eqref{eq:boundfinelista1}-\eqref{eq:boundfinelista2} into \eqref{tosplit}, then dividing by $\big\|A^{\mathbf{2}}\|_{\mathrm{op}}\big\|B^{\mathbf{2}}\|_{\mathrm{op}}$, and taking the supremum in the l.h.s., finally yields
  \begin{equation}\label{thestartingpoint}
\begin{split}
    \mathcal{F}_{\substack{m_1, m_2 \\ n_1, n_2}}(t) \;&\leqslant\;   \frac{4 n_1 m_1 \|V_1\|_{L^\infty} t }{N_1}  + \frac{ 4 n_2 m_2 \|V_2\|_{L^\infty} t}{N_2}  \\
    & \;\; + \frac{4 n_1 m_2 \|V_{12}\|_{L^\infty}t}{N_1+N_2}  + \frac{4 n_2 m_1 \|V_{12}\|_{L^\infty}t}{N_1+N_2}   \\
    &\;\; + 2 n_1 \|V_1\|_{L^\infty}\int_0^t \ud s\, \mathcal{F}_{\substack{m_1, m_2 \\ 2, 0}}(s) + 2 n_2  \|V_2\|_{L^\infty}\int_0^t \ud s\, \mathcal{F}_{\substack{m_1, m_2 \\ 0, 2}}(s) \\
    & \;\;  + 2 (n_1+n_2) (1-c)\|\widehat{V}_{12}\|_{L^1}  \int_0^t  \ud s\, \mathcal{F}_{\substack{m_1, m_2 \\ 1, 1}}(s)\,.
    \end{split}
\end{equation}

 Estimate \eqref{thestartingpoint} has the structure of a multi-component Gr\o{}nwall-type inequality and is thus suited to provide an explicit bound on $ \mathcal{F}_{\substack{m_1, m_2 \\ n_1, n_2}}(t)$. Replacing in it the various potential norms with their common majorant $\mathcal{W}:=\frac{1}{12}\mathcal{V}$ gives
 \begin{equation}\label{eq:Fest2}
  \begin{split}
   &\mathcal{F}_{\substack{m_1, m_2 \\ n_1, n_2}}(t) \;\leqslant\; 4\,\mathcal{W}\,t\,\Big(\frac{m_1 n_1}{N_1}+\frac{m_2 n_2}{N_2}+\frac{m_1 n_2+m_2 n_1}{N_1+N_2}\Big) \\
   &\qquad + 2\,\mathcal{W}\,(n_1+n_2)\int_0^t\ud s\,\big(\mathcal{F}_{\substack{m_1, m_2 \\ 2, 0}}(s)+\mathcal{F}_{\substack{m_1, m_2 \\  0,2}}(s)+\mathcal{F}_{\substack{m_1, m_2 \\ 1,1}}(s)\big)\,.
  \end{split}
 \end{equation}
 Specialising \eqref{eq:Fest2} for $(n_1,n_2)=(2,0)$, $(0,2)$, and $(1,1)$, and adding up the three inequalities thus obtained, gives
 \begin{equation*}
  \begin{split}
   &\mathcal{F}_{\substack{m_1, m_2 \\ 2, 0}}(t)+\mathcal{F}_{\substack{m_1, m_2 \\  0,2}}(t)+\mathcal{F}_{\substack{m_1, m_2 \\ 1,1}}(t) \;\leqslant\;12\,\mathcal{W}\, t\,\Big(\frac{m_1}{N_1}+\frac{m_2}{N_2}+\frac{m_1+m_2}{N_1+N_2}\Big) \\
   &\qquad +12\,\mathcal{W}\int_0^t\ud s\,\big(\mathcal{F}_{\substack{m_1, m_2 \\ 2, 0}}(s)+\mathcal{F}_{\substack{m_1, m_2 \\  0,2}}(s)+\mathcal{F}_{\substack{m_1, m_2 \\ 1,1}}(s)\big) \\
   & \leqslant\;\frac{\,24\,\mathcal{W}\,t}{c\,N}(m_1+m_2)+12\,\mathcal{W}\int_0^t\ud s\,\big(\mathcal{F}_{\substack{m_1, m_2 \\ 2, 0}}(s)+\mathcal{F}_{\substack{m_1, m_2 \\  0,2}}(s)+\mathcal{F}_{\substack{m_1, m_2 \\ 1,1}}(s)\big)\,,
  \end{split}
 \end{equation*}
 where we further used the bound $\big(\frac{m_1}{N_1}+\frac{m_2}{N_2}+\frac{m_1+m_2}{N_1+N_2}\big)\leqslant\frac{2}{cN}(m_1+m_2)$ in terms of $N=N_1+N_2$ and $c=\min\{\frac{N_1}{N},\frac{N_2}{N}\}$. A standard Gr\o{}nwall-Gollwitzer inequality (see, e.g., \cite[Theorem 1.3.2]{Pachpatte-ineq}) then yields
 \begin{equation}\label{eq:f200211}
  \mathcal{F}_{\substack{m_1, m_2 \\ 2, 0}}(t)+\mathcal{F}_{\substack{m_1, m_2 \\  0,2}}(t)+\mathcal{F}_{\substack{m_1, m_2 \\ 1,1}}(t) \;\leqslant\;\frac{\,2(m_1+m_2)}{c\,N}(\text{e}^{12\mathcal{W} t}-1)\,.
 \end{equation}
  Plugging \eqref{eq:f200211} back into \eqref{eq:Fest2} and estimating
  \[
   \frac{m_1 n_1}{N_1}+\frac{m_2 n_2}{N_2}+\frac{m_1 n_2+m_2 n_1}{N_1+N_2}\;\leqslant\;\frac{(n_1+n_2)(m_1+m_2)}{c\,N}
  \]
  finally produces
  \[
   \mathcal{F}_{\substack{m_1, m_2 \\ n_1, n_2}}(t) \;\leqslant\;\frac{\,(m_1+m_2)(n_1+n_2)\,}{3\,c\,N}(\text{e}^{12\mathcal{W} t}-1)
  \]
 which has precisely the form of \eqref{eq:whatwewant}.

 The proof of  Theorem \ref{thm:LR} is thus completed.

 \section{Correlation estimates}\label{sec:corr}

 The proof of Theorem \ref{thm:main1} is consists of a convenient expansion of the considered correlation function and manipulation of the various outcoming terms in order to produce in each of them (expectations of) commutators of the form of those controlled by the Lieb-Robinson bounds of Theorem \ref{thm:LR}. This procedure will be discussed in separate steps.

 \subsection{Insertion of projections}~
 
  In view of the short-hand \eqref{shorthand2}, and the further short-hand
  \begin{equation}\label{eq:shAt}
   \begin{split}
    A^{\mathbf{1}}_t \; \equiv \; \text{e}^{itH_{N_1, N_2}}\,A^{\mathbf{1}} \,\text{e}^{-itH_{N_1, N_2}}\,, \\
    A^{\mathbf{2}}_t \; \equiv \; \text{e}^{itH_{N_1, N_2}}\,A^{\mathbf{2}} \,\text{e}^{-itH_{N_1, N_2}}\,, \\
    B^{\mathbf{1}}_t \; \equiv \; \text{e}^{itH_{N_1, N_2}}\,B^{\mathbf{1}} \,\text{e}^{-itH_{N_1, N_2}}\,, \\
    B^{\mathbf{2}}_t \; \equiv \; \text{e}^{itH_{N_1, N_2}}\,B^{\mathbf{2}} \,\text{e}^{-itH_{N_1, N_2}}\,,
   \end{split}
  \end{equation}
  and using the fact that $\Psi_{N_1,N_2,t}=\text{e}^{-\ii t H_{N_1,N_2}} (u^{\otimes N_1}\otimes v^{\otimes N_2})$,   
  the correlation function considered in \eqref{claimthgrowth2} is re-written as
  \begin{equation}
\begin{split}
    \big\langle &A^{\mathbf{2}}  B^{\mathbf{2}} A^{\mathbf{1}} B^{\mathbf{1}} \big\rangle_{N_1, N_2,t}-\big\langle A^{\mathbf{2}} B^{\mathbf{2}} \big\rangle_{N_1, N_2,t} \big\langle A^{\mathbf{1}} B^{\mathbf{1}} \big\rangle_{N_1, N_2, t} \\
    &\qquad =\;\big\langle A^{\mathbf{2}}_t  B^{\mathbf{2}}_t A^{\mathbf{1}}_t B^{\mathbf{1}}_t \big\rangle_{N_1, N_2}-\big\langle A^{\mathbf{2}}_t B^{\mathbf{2}}_t \big\rangle_{N_1, N_2} \big\langle A^{\mathbf{1}}_t B^{\mathbf{1}}_t \big\rangle_{N_1, N_2}\,.
\end{split}
\end{equation}

  We now plan to produce inside the above expressions orthogonal projections like $p_u:=|u\rangle\langle u|$ and $p_v:=|v\rangle\langle v|$, so as to monitor the content of a $u$-particle or $v$-particle in each considered variable. To this aim we insert the identity on the $N_1$-body and the $N_2$-body space in the form of the expansions
  \begin{equation*}
   \begin{split}
    \mathbbm{1}_{N_1}\,&=\,(p_u+(\mathbbm{1}-p_u))^{\otimes N_1}\,=\,p_u^{\otimes N_1} +\sum_{j_1=1}^{N_1} p_u^{\otimes (j_1-1)} \otimes (\mathbbm{1}-p_u) \otimes \mathbbm{1}^{\otimes (N_1-j_1)}\,, \\
    \mathbbm{1}_{N_2}\,&=\,(p_v+(\mathbbm{1}-p_v))^{\otimes N_2}\,=\,p_v^{\otimes N_2} +\sum_{j_2=1}^{N_2} p_v^{\otimes (j_2-1)} \otimes (\mathbbm{1}-p_v) \otimes \mathbbm{1}^{\otimes (N_2-j_2)}
   \end{split}
  \end{equation*}
  that can be re-written as
  \begin{equation}\label{identities2}
   \begin{split}
    \mathbbm{1}_{N_1}\,&=\,p_u^{\otimes N_1} +\sum_{j_1=1}^{N_1} p_u^{[1,\dots,j_1-1]}q_u^{[j_1]}\,, \\
    \mathbbm{1}_{N_2}\,&=\,p_v^{\otimes N_2} +\sum_{j_2=1}^{N_2} p_v^{[1, \dots, j_2-1]} q_v^{[j_2]}
   \end{split}
  \end{equation}
  by means of the short-hand
\begin{equation}\label{defproj}
    \begin{split}
      q_u^{[j_1]} \;&:=\; \mathbbm{1}^{\otimes (j-1)} \otimes (\mathbbm{1}-p_u) \otimes \mathbbm{1}^{\otimes (N_1-j_1)}\,,\\
     q_v^{[j_2]} \;&:=\;\mathbbm{1}^{\otimes (j_2-1)} \otimes (\mathbbm{1}-p_v) \otimes \mathbbm{1}^{\otimes (N_2-j_2)}\,,\\
     p_u^{[1, \dots, j_1-1]} \;&:=\; p_u^{\otimes (j_1-1)} \otimes \mathbbm{1}^{\otimes (N_1-j_1+1)}\,,\\
     p_v^{[1, \dots, j_2-1]}\;&:=\; p_v^{\otimes (j_2-1)} \otimes  \mathbbm{1}^{\otimes (N_2-j_2+1)}\,.
\end{split}
\end{equation}
 Indeed, for instance, the $j_1$-th element of the sum in the first identity \eqref{identities2} is
\begin{equation*}
\begin{split}
        p_u^{[1, \dots, j_1-1]}  q_u^{[j_1]} \; &=\;  (p_u^{\otimes (j_1-1)} \otimes \mathbbm{1}^{\otimes (N_1-j_1+1)}) ( \mathbbm{1}^{\otimes (j-1)} \otimes (\mathbbm{1}-p_u) \otimes 1^{\otimes (N_1-j_1)}) \\
       & =\; p_u^{\otimes (j_1-1)} \otimes (\mathbbm{1}-p_u) \otimes \mathbbm{1}^{\otimes (N_1-j_1)}\,.
\end{split}
\end{equation*}
  All operators \eqref{defproj} are orthogonal projections and have therefore unit norm. The overall identity $\mathbbm{1}_{N_1}\otimes\mathbbm{1}_{N_2}$ on $\mathcal{H}_{N_1,N_2}$ can then be expressed as
 \begin{equation}\label{tu}
\begin{split}
     \mathbbm{1}_{N_1,N_2} \;&=\;  p_u^{\otimes N_1}\otimes  p_v^{\otimes N_2} + \biggl(\sum_{j_1=1}^{N_1} p_u^{[1, \dots, j_1-1]}  q_u^{[j_1]}\biggr) \otimes \biggl( \sum_{j_2=1}^{N_2} p_v^{[1, \dots, j_2-1]}  q_v^{[j_2]} \biggr) \\
     & \quad + p_u^{\otimes N_1} \otimes \biggl( \sum_{j_2=1}^{N_2} p_v^{[1, \dots, j_2-1]} q_v^{[j_2]}\biggr) +  \biggl(\sum_{j_1=1}^{N_1} p_u^{[1, \dots, j_1-1]}  q_u^{[j_1]}\biggr) \otimes p_v^{\otimes N_2}\,,
\end{split}
\end{equation}
 the above tensor products referring to the product structure \eqref{eq:H12-tensorproduct}.

  Upon plugging \eqref{tu} into $\big\langle A^{\mathbf{2}}_t  B^{\mathbf{2}}_t \,\mathbbm{1}_{N_1,N_2} A^{\mathbf{1}}_t B^{\mathbf{1}}_t \big\rangle_{N_1, N_2}$, and observing that 
  \begin{equation*}
\begin{split}
    &\langle A^{\mathbf{2}}_t\, B^{\mathbf{2}}_t\,\, p_u^{\otimes N_1}\otimes  p_v^{\otimes N_2} A^{\mathbf{1}}_t\, B^{\mathbf{1}}_t \rangle_{N_1, N_2}\\
    &\qquad=\;  \langle  A^{\mathbf{2}}_t\, B^{\mathbf{2}}_t | u^{\otimes N_1} \otimes v^{\otimes N_2} \rangle \langle  u^{\otimes N_1} \otimes v^{\otimes N_2} | A^{\mathbf{1}}_t\, B^{\mathbf{1}}_t  \rangle_{N_1, N_2}\\
    &\qquad=\; \langle  A^{\mathbf{2}}_t\, B^{\mathbf{2}}_t\rangle_{N_1, N_2}\langle A^{\mathbf{1}}_t\, B^{\mathbf{1}}_t  \rangle_{N_1, N_2}\,,
\end{split}
\end{equation*}
  we obtain
  \begin{equation}\label{sta}
   \begin{split}
    & \big\langle A^{\mathbf{2}}_t  B^{\mathbf{2}}_t A^{\mathbf{1}}_t B^{\mathbf{1}}_t \big\rangle_{N_1, N_2}-\big\langle A^{\mathbf{2}}_t B^{\mathbf{2}}_t \big\rangle_{N_1, N_2} \big\langle A^{\mathbf{1}}_t B^{\mathbf{1}}_t \big\rangle_{N_1, N_2} \\
    &\qquad =\; \mathcal{P}_{j_1}^{N_1}+\mathcal{Q}_{j_2}^{N_2}+\mathcal{R}_{j_1,j_2}^{N_1,N_2}
   \end{split}
  \end{equation}
  with
\begin{equation}\label{notationabbeta}
\begin{split}
   \mathcal{P}_{j_1}^{N_1} \;&:=\; \sum_{j_1=1}^{N_1} \langle  A^{\mathbf{2}}_t\, B^{\mathbf{1}}_t\, p_u^{[1, \dots, j_1-1]}  q_u^{[j_1]}\,\otimes p_v^{\otimes N_2}\, \, A^{\mathbf{1}}_t\, B^{\mathbf{1}}_t \rangle_{N_1, N_2}\,, \\
   \mathcal{Q}_{j_2}^{N_2} \;&:=\; \sum_{j_2=1}^{N_2}  \langle  A^{\mathbf{2}}_t\, B^{\mathbf{1}}_t\, \,\, p_u^{\otimes N_1} \otimes  p_v^{[1, \dots, j_2-1]} q_v^{[j_2]}\,A^{\mathbf{1}}_t\, B^{\mathbf{1}}_t \rangle_{N_1, N_2}\,,  \\
  \mathcal{R}_{j_1,j_2}^{N_1,N_2} \;&:=\;\sum_{j_1=1}^{N_1} \sum_{j_2=1}^{N_2}  \langle  A^{\mathbf{2}}_t\, B^{\mathbf{1}}_t \,\,p_u^{[1, \dots, j_1-1]}  q_u^{[j_1]} \otimes p_v^{[1, \dots, j_2-1]}  q_v^{[j_2]}\,\,A^{\mathbf{1}}_t\, B^{\mathbf{1}}_t \rangle_{N_1, N_2}\,.
\end{split}
\end{equation}
  In the following we shall control the three contributions in the r.h.s.~of \eqref{sta}.

 \subsection{Control of the $\mathcal{P}_{j_1}^{N_1}$-terms and $\mathcal{Q}_{j_2}^{N_2}$-terms}~

  It suffices to estimate only one type of contributions, as the other type is obtained by merely exchanging $j_1\leftrightarrow j_2$ and $N_1\leftrightarrow N_2$.

  It is natural to split
  \begin{equation}\label{secondtermsplit}
    \begin{split}
      \mathcal{Q}_{j_2}^{N_2} \;&= \;\sum_{j_2=1}^{n} \langle  A^{\mathbf{2}}_t\, B^{\mathbf{2}}_t\, \,\, p_u^{\otimes N_1} \otimes  p_v^{[1, \dots, j_2-1]} q_v^{[j_2]}\, A^{\mathbf{1}}_t\, B^{\mathbf{1}}_t \rangle_{N_1, N_2}   \\
        &\qquad +\sum_{j_2=n+1}^{n+m} \langle  A^{\mathbf{2}}_t\, B^{\mathbf{2}}_t\, \, p_u^{\otimes N_1} \otimes  p_v^{[1, \dots, j_2-1]} q_v^{[j_2]}\, A^{\mathbf{1}}_t\, B^{\mathbf{1}}_t \rangle_{N_1, N_2}  \\
        &\qquad \qquad  +\sum_{j_2=n+m+1}^{N_2} \langle  A^{\mathbf{2}}_t\, B^{\mathbf{2}}_t\, \, p_u^{\otimes N_1} \otimes  p_v^{[1, \dots, j_2-1]} q_v^{[j_2]}\, A^{\mathbf{1}}_t\, B^{\mathbf{1}}_t \rangle_{N_1, N_2}  \,.
    \end{split} 
\end{equation}
  In each of the terms above it is convenient to produce commutators whose norm can be then estimated by means of the Lieb-Robinson bounds established in Theorem \ref{thm:LR}. The typical form of such bounds required here is
  \begin{equation}\label{eq:ourLR}
  \begin{split}
      \sup_{ \substack{  A_1\in\mathcal{B}'_{n}, \,A_2\in\mathcal{B}'_{m} \\  B_2\in\mathcal{B}'_{n}, \,B_2\in\mathcal{B}'_{m}}} \frac{\;\big\| \big[ A^{\mathbf{2}}B^{\mathbf{2}} ,A_t^{\mathbf{1}}B_t^{\mathbf{1}}\big] \big\|_{\mathrm{op}}\,}{\|A^{\mathbf{1}}\|_{\mathrm{op}}\|A^{\mathbf{2}}\|_{\mathrm{op}}\|B^{\mathbf{1}}\|_{\mathrm{op}}\|B^{\mathbf{2}}\|_{\mathrm{op}}}   \;\leqslant\;\frac{4mn}{3 c N}(\text{e}^{\mathcal{V} t}-1)
  \end{split}
  \end{equation}
  obtained by \eqref{eq:LRbound-general} by choosing $m_1=m_2=m$ and $n_1=n_2=n$, replacing $12\mathcal{W}=\mathcal{V}$, and using the short-hand \eqref{eq:shAt}.

  Let us stress that for the applicability of \eqref{eq:ourLR} the particle indices with respect to which one builds, according to \eqref{shorthand2}, the operators  \emph{at time zero} $A^{\mathbf{1}},B^{\mathbf{1}}$ on the one hand and $A^{\mathbf{2}},B^{\mathbf{2}}$ on the other \emph{must be disjoint}.

  We thus re-write \eqref{secondtermsplit} as
  \begin{equation}\label{secondtermsplit2}
    \begin{split}
        \mathcal{Q}_{j_2}^{N_2} \;&= \sum_{j_2=1}^{n} \langle\,[A^{\mathbf{2}}_t\, B^{\mathbf{2}}_t,q_v^{[j_2]}]\,\, p_u^{\otimes N_1} \otimes  p_v^{[1, \dots, j_2-1]}A^{\mathbf{1}}_t\, B^{\mathbf{1}}_t \rangle_{N_1, N_2}  \\
        &\quad +\sum_{j_2=n+1}^{n+m} \langle A^{\mathbf{2}}_t\, B^{\mathbf{2}}_t\,\, p_u^{\otimes N_1} \otimes  p_v^{[1, \dots, j_2-1]}\, [q_v^{[j_2]},\,A^{\mathbf{1}}_t\, B^{\mathbf{1}}_t] \,\rangle_{N_1, N_2} \\
        &\quad +\sum_{j_2=n+m+1}^{N_2} \langle \,[ A^{\mathbf{2}}_t\, B^{\mathbf{2}}_t,\, q_v^{[j_2]}]\, p_u^{\otimes N_1} \otimes  p_v^{[1, \dots, j_2-1]}\,[q_v^{[j_2]},\,A^{\mathbf{1}}_t\, B^{\mathbf{1}}_t]\, \rangle_{N_1, N_2} \,.
    \end{split} 
\end{equation}
  For the justification of \eqref{secondtermsplit2}, observe that in the first line, the commutator
  \[
     [ A^{\mathbf{2}}_t\, B^{\mathbf{2}}_t, q_v^{[j_2]}] \;=\; A^{\mathbf{2}}_t\, B^{\mathbf{2}}_t q_v^{[j_2]} - q_v^{[j_2]}A^{\mathbf{2}}_t\, B^{\mathbf{2}}_t
  \]
  only contributes with its first summand, for
  \[
   q_v^{[j_2]} \Psi_{N_1, N_2} \;=\;  q_v^{[j_2]} (u^{\otimes N_1} \otimes v^{\otimes N_2}) \;=\;0\,,
  \]
  so this is the same as the first line in \eqref{secondtermsplit}. The second and third line in \eqref{secondtermsplit2} are validated analogously, in particular in the third line the orthogonal projection property $ (q_v^{[j_2]})^2 = q_v^{[j_2]}$ was also used.

  The commutators in \eqref{secondtermsplit2} all have the appropriate content of particles needed to apply the Lieb-Robinson bounds \eqref{eq:ourLR}. For example, in the commutator $[q_v^{[j_2]},A^{\mathbf{2}}_t\, B^{\mathbf{2}}_t]$ in the first line, the index $j_2$ belongs to $\{1,\dots,n\}$ and is therefore \emph{disjoint} from the set of indices $\{n+1,\dots,n+m\}$ labelling both $A^{\mathbf{2}}$ and $B^{\mathbf{2}}$ \emph{at time zero}. An analogous disjointness is immediately checked for all other commutators, depending on where $j_2$ runs over.

  Thus, \eqref{eq:ourLR} yields
  \begin{equation*}
   \begin{split}
    \big\|[A^{\mathbf{2}}_t\, B^{\mathbf{2}}_t,q_v^{[j_2]}] \big\|_{\mathrm{op}} \;&=\;\big\| [q_v^{[j_2]}, A^{\mathbf{2}}_t\, B^{\mathbf{2}}_t] \big\|_{\mathrm{op}} \\
    &\leqslant\; \frac{4 m \|A^{\mathbf{2}}\|_{\mathrm{op}} \|B^{\mathbf{2}}\|_{\mathrm{op}} }{ 3 c N}(\text{e}^{ \mathcal{V}t}-1)\qquad \textrm{when }j_2\in\{1,\dots,n\}
   \end{split}
  \end{equation*}
 and the first line in \eqref{secondtermsplit2} is therefore estimated as
   \begin{equation}\label{eq:Qline1}
    \begin{split}
     &\bigg| \sum_{j_2=1}^{n} \langle\,[A^{\mathbf{2}}_t\, B^{\mathbf{2}}_t,q_v^{[j_2]}]\,\, p_u^{\otimes N_1} \otimes  p_v^{[1, \dots, j_2-1]}A^{\mathbf{1}}_t\, B^{\mathbf{1}}_t \rangle_{N_1, N_2} \bigg| \\
     &\quad\leqslant\;\|A^{\mathbf{1}}_t\|_{\mathrm{op}} \|B^{\mathbf{1}}_t\|_{\mathrm{op}}\sum_{j_2=1}^{n}\big\|[A^{\mathbf{2}}_t\, B^{\mathbf{2}}_t,q_v^{[j_2]}] \big\|_{\mathrm{op}}\big\|p_u^{\otimes N_1} \otimes  p_v^{[1, \dots, j_2-1]}\big\|_{\mathrm{op}} \\
     &\quad\leqslant\;\frac{4mn\mathcal{L}}{3cN}(\text{e}^{\mathcal{V}t}-1)\,,
    \end{split}
   \end{equation}
   having set for convenience
   \begin{equation}\label{eq:Lshorthand}
    \begin{split}
            \mathcal{L}\;:=&\;\|A^{\mathbf{1}}_t\|_{\mathrm{op}} \|A^{\mathbf{2}}_t\|_{\mathrm{op}}\|B^{\mathbf{1}}_t\|_{\mathrm{op}} \|B^{\mathbf{2}}_t\|_{\mathrm{op}} \\
            =&\;\|A_1\|_{\mathrm{op}} \|A_2\|_{\mathrm{op}}\|B_1\|_{\mathrm{op}} \|B_2\|_{\mathrm{op}}
    \end{split}
   \end{equation}
  and having used the fact that orthogonal projections have unit norm.   
  Analogously, for the second line in \eqref{secondtermsplit2} we find from \eqref{eq:ourLR}
  \[
   \big\| [q_v^{[j_2]}, A^{\mathbf{1}}_t\, B^{\mathbf{1}}_t] \big\|\;\leqslant\; \frac{4 n \|A^{\mathbf{1}}\|_{\mathrm{op}} \|B^{\mathbf{1}}\|_{\mathrm{op}} }{ 3 c N}(\text{e}^{ \mathcal{V}t}-1)\,,\qquad j_2\in\{n+1,\dots,n+m\}\,,
  \]
  whence  
     \begin{equation}\label{eq:Qline2}
    \begin{split}
     &\bigg| \sum_{j_2=n+1}^{n+m} \langle A^{\mathbf{2}}_t\, B^{\mathbf{2}}_t\,\, p_u^{\otimes N_1} \otimes  p_v^{[1, \dots, j_2-1]} \,[q_v^{[j_2]},\,A^{\mathbf{1}}_t\, B^{\mathbf{1}}_t]\, \rangle_{N_1, N_2} \bigg| \\
     &\quad\leqslant\;\|A^{\mathbf{2}}_t\|_{\mathrm{op}} \|B^{\mathbf{2}}_t\|_{\mathrm{op}}\sum_{j_2=n+1}^{n+m}\big\| [q_v^{[j_2]}, A^{\mathbf{1}}_t\, B^{\mathbf{1}}_t] \big\|_{\mathrm{op}} \;\leqslant\;\frac{4mn\mathcal{L}}{3cN}(\text{e}^{\mathcal{V}t}-1)\,.
    \end{split}
   \end{equation}
   Last, for the third line in \eqref{secondtermsplit2} we find from \eqref{eq:ourLR}
  \[
   \begin{split}
    \big\| [q_v^{[j_2]}, A^{\mathbf{1}}_t\, B^{\mathbf{1}}_t] \big\|_{\mathrm{op}}\;&\leqslant\; \frac{4 n \|A^{\mathbf{1}}\|_{\mathrm{op}} \|B^{\mathbf{1}}\|_{\mathrm{op}} }{ 3 c N}(\text{e}^{ \mathcal{V}t}-1)\,, \\
    \big\| [q_v^{[j_2]}, A^{\mathbf{2}}_t\, B^{\mathbf{2}}_t] \big\|_{\mathrm{op}}\;&\leqslant\; \frac{4 m \|A^{\mathbf{2}}\|_{\mathrm{op}} \|B^{\mathbf{2}}\|_{\mathrm{op}} }{ 3 c N}(\text{e}^{ \mathcal{V}t}-1)\,,
   \end{split} \qquad j_2\in\{ n+m+1,\dots N_2\}\,,
  \]
  whence
  \begin{equation}\label{eq:Qline3}
   \begin{split}
    &\bigg|\sum_{j_2=n+m+1}^{N_2} \langle \,[ A^{\mathbf{2}}_t\, B^{\mathbf{2}}_t,\, q_v^{[j_2]}]\, p_u^{\otimes N_1} \otimes  p_v^{[1, \dots, j_2-1]}\,[q_v^{[j_2]},\,A^{\mathbf{1}}_t\, B^{\mathbf{1}}_t]\, \rangle_{N_1, N_2}\bigg| \\
    &\quad \leqslant\;(N_2-n-m)\,\frac{16mn\mathcal{L}}{9 c^2 N^2}(\text{e}^{ \mathcal{V}t}-1)^2\;\leqslant\;\frac{16(1-c)mn\mathcal{L}}{9 c^2 N}(\text{e}^{ \mathcal{V}t}-1)^2\,,
   \end{split}
  \end{equation}
  having used $\frac{N_2-n-m}{N}\leqslant 1-c$. The latter estimate also shows why for an efficient control of the third line in \eqref{secondtermsplit} \emph{two commutators} had to be produced therein, instead of just one as for the previous lines: that sum has a large $O(N_2)$ number of summands, and a single commutator makes each such term $O(N^{-1})$-small, thereby yielding a non-vanishing $O(1)$-quantity.

  Combining \eqref{secondtermsplit2}, \eqref{eq:Qline1}, \eqref{eq:Qline2}, and \eqref{eq:Qline3} together finally gives
  \begin{equation}\label{bounbj2}
\big|\mathcal{Q}_{j_2}^{N_2}\big|\; \leqslant\; \frac{8 m n \mathcal{L}}{3 c N}  (\text{e}^{ \mathcal{V}t}-1) \Big(1+\frac{2(1-c)}{3 c} (\text{e}^{\mathcal{V}t}-1)\Big)\,.
\end{equation}
  As argued already, the reasoning for $\mathcal{P}_{j_1}^{N_1}$ is analogous, thus
     \begin{equation}\label{bounbj1}
\big|\mathcal{P}_{j_1}^{N_1}\big|\; \leqslant\; \frac{8 m n \mathcal{L}}{3 c N}  (\text{e}^{\mathcal{V}t}-1) \Big(1+\frac{2(1-c)}{3 c} (\text{e}^{\mathcal{V}t}-1)\Big)\,.
\end{equation}

  \subsection{Control of the terms of $O(1)$ and $O(N)$ size in $\mathcal{R}_{j_1,j_2}^{N_1,N_2}$}~
  
  One sees from \eqref{notationabbeta} that the part $\mathcal{R}_{j_1,j_2}^{N_1,N_2}$ of the correlation function \eqref{sta} has formal size $O(N^2)$, meant as the number of terms of the double sum therein. Let us separate the sub-leading contribution coming from the sole portion of the first $n$ and the next $m$ A-particles or B-particles, which shall be estimated now, from the genuinely $O(N^2)$-large part coming from pairings between such `internal' particles and all the remaining `external' ones, which is the object of the next subsection.

  We thus split
  \begin{equation}\label{eq:Rsplit}
   \mathcal{R}_{j_1,j_2}^{N_1,N_2}\;:=\;\mathrm{(I)}+\mathrm{(II)}+\mathrm{(III)}+\mathrm{(IV)}
  \end{equation}
  with
  \begin{equation}
   \mathcal{E}_{j_1, j_2}\;:=\;  \langle A^{\mathbf{2}}_t\, B^{\mathbf{2}}_t \,\,p_u^{[1, \dots, j_1-1]} q_u^{[j_1]} \otimes p_v^{[1, \dots, j_2-1]} q_v^{[j_2]}\,\, A^{\mathbf{1}}_t\, B^{\mathbf{1}}_t \rangle_{N_1, N_2}
  \end{equation}
  and
  \begin{equation}
   \begin{split}
    \mathrm{(I)}\;&:=\;\sum_{j_1=1}^{n}\sum_{j_2=1}^{n}\mathcal{E}_{j_1, j_2} + \sum_{j_1=1}^{n}\sum_{j_2=n+1}^{n+m}\mathcal{E}_{j_1, j_2} \\
    & \qquad \qquad \qquad +\sum_{j_1=n+1}^{n+m}\sum_{j_2=1}^{n}\mathcal{E}_{j_1, j_2}+\sum_{j_1=n+1}^{n+m}\sum_{j_2=n+1}^{n+m}\mathcal{E}_{j_1, j_2}\,, \\
    \mathrm{(II)}\;&:=\;\sum_{j_1=n+m+1}^{N_1}\sum_{j_2=1}^{n}\mathcal{E}_{j_1, j_2} +\sum_{j_1=n+m+1}^{N_1}\sum_{j_2=n+1}^{n+m}\mathcal{E}_{j_1, j_2}\,, \\
    \mathrm{(III)}\;&:=\;\sum_{j_1 = 1}^{n}\sum_{j_2=n+m+1}^{N_2}\mathcal{E}_{j_1, j_2} +\sum_{j_1 = n+1}^{n+m}\sum_{j_2=n+m+1}^{N_2}\mathcal{E}_{j_1, j_2}\,, \\
    \mathrm{(IV)}\;&:=\;\sum_{j_1 = n+m+1}^{N_1}\sum_{j_2=n+m+1}^{N_2}\mathcal{E}_{j_1, j_2}
   \end{split}
  \end{equation}
  and consider here the contributions $\mathrm{(I)}$, of formal size $O(1)$, and  $\mathrm{(II)}$, $\mathrm{(III)}$, of formal size $O(N)$.

  In the first double sum in $\mathrm{(I)}$,
  \[
    \mathcal{E}_{j_1, j_2}\;=\; \langle \,[A^{\mathbf{2}}_t\, B^{\mathbf{2}}_t,  q_u^{[j_1]} \otimes q_v^{[j_2]}]\,\,  p_u^{[1, \dots, j_1-1]}  \otimes  p_v^{[1, \dots, j_2-1]}\,\, A^{\mathbf{1}}_t\, B^{\mathbf{1}}_t \rangle_{N_1, N_2}
  \]
  because, as usual, the term $(q_u^{[j_1]} \otimes q_v^{[j_2]})A^{\mathbf{2}}_t\, B^{\mathbf{2}}_t$ of the commutator does not contribute in the expectation, as it annihilates $\Psi_{N_1,N_2}=u^{\otimes N_1}\otimes v^{\otimes N_2}$ when acting onto it on the left in the expectation, and moreover \eqref{eq:ourLR} in this case implies
  \begin{equation*}
    \big\| [q_u^{[j_1]} \otimes q_v^{[j_2]}, A^{\mathbf{2}}_t\, B^{\mathbf{2}}_t] \big\|_{\mathrm{op}} \;\leqslant\; \frac{4 m \|A^{\mathbf{2}}\|_{\mathrm{op}} \|B^{\mathbf{2}}\|_{\mathrm{op}}}{3 c N} (\text{e}^{\mathcal{V}t}-1)\,,\qquad j_1,j_2\in\{1,\dots,n\}\,,
\end{equation*}
 whence, using also \eqref{eq:Lshorthand},
 \begin{equation}\label{eq:I1}
  \begin{split}
   \bigg| \sum_{j_1=1}^{n}\sum_{j_2=1}^{n}\mathcal{E}_{j_1, j_2} \bigg|\;&\leqslant\; \| A^{\mathbf{1}}_t\|_{\mathrm{op}} \|B^{\mathbf{1}}_t\|_{\mathrm{op}} \sum_{j_1=1}^{n} \sum_{j_2=1}^{n} \big\| [q_u^{[j_1]} \otimes q_v^{[j_2]}, A^{\mathbf{2}}_t\, B^{\mathbf{2}}_t] \big\|_{\mathrm{op}} \\
   &\leqslant\;\frac{\,4mn^2\mathcal{L}}{3cN}(\text{e}^{\mathcal{V}t}-1)\,.
  \end{split}
 \end{equation}

  In the second double sum in $\mathrm{(I)}$,
  \[
    \mathcal{E}_{j_1, j_2}\;=\; \langle \,[A^{\mathbf{2}}_t\, B^{\mathbf{2}}_t, q_u^{[j_1]}] \,\, p_u^{[1, \dots, j_1-1]}  \otimes  p_v^{[1, \dots, j_2-1]}\,\, [q_v^{[j_2]}, \,A^{\mathbf{1}}_t\, B^{\mathbf{1}}_t]\, \rangle_{N_1, N_2} \,,
  \]
 the commutators being produced again because $q_u^{[j_1]}$ and $q_v^{[j_2]}$ annihilate $\Psi_{N_1,N_2}$ on the two different sides of the expectation, and moreover \eqref{eq:ourLR} implies
  \[
   \begin{split}
     \big\| [q_u^{[j_1]}, A^{\mathbf{2}}_t\, B^{\mathbf{2}}_t] \big\|_{\mathrm{op}} \;&\leqslant\; \frac{4 m \|A^{\mathbf{2}}\|_{\mathrm{op}} \|B^{\mathbf{2}}\|_{\mathrm{op}}}{3 c N} (\text{e}^{\mathcal{V}t}-1)\,,\qquad j_1\in\{1,\dots,n\}\,, \\
     \big\| [q_v^{[j_2]}, A^{\mathbf{1}}_t\, B^{\mathbf{1}}_t] \big\|_{\mathrm{op}} \;&\leqslant\; \frac{4 n \|A^{\mathbf{1}}\|_{\mathrm{op}} \|B^{\mathbf{1}}\|_{\mathrm{op}}}{3 c N} (\text{e}^{\mathcal{V}t}-1)\,,\qquad j_2\in\{n+1,\dots,n+m\}\,,
   \end{split}
  \]
  whence
\begin{equation}\label{eq:I2}
     \bigg| \sum_{j_1=1}^{n}\sum_{j_2=n+1}^{n+m}\mathcal{E}_{j_1, j_2} \bigg|\;\leqslant\; \frac{\,16\,m^2n^2\mathcal{L}}{9c^2N^2}(\text{e}^{\mathcal{V}t}-1)\,.
 \end{equation}
  As the third double sum in $\mathrm{(I)}$ is obtained from the second by merely exchanging the indices of the A-particles and the B-particles, we also find
\begin{equation}\label{eq:I3}
     \bigg| \sum_{j_1=n+1}^{n+m}\sum_{j_2=1}^{n}\mathcal{E}_{j_1, j_2} \bigg|\;\leqslant\; \frac{\,16\,m^2n^2\mathcal{L}}{9c^2N^2}(\text{e}^{\mathcal{V}t}-1)\,.
 \end{equation}

 For the fourth double sum in $\mathrm{(I)}$ we have, in analogy to the first double sum,
 \[
    \mathcal{E}_{j_1, j_2}\;=\;\langle \,A^{\mathbf{2}}_t\, B^{\mathbf{2}}_t \,\,  p_u^{[1, \dots, j_1-1]}  \otimes  p_v^{[1, \dots, j_2-1]}\,\,[q_u^{[j_1]}\otimes q_v^{[j_2]},\,A^{\mathbf{1}}_t\, B^{\mathbf{1}}_t]\, \rangle_{N_1, N_2}
  \]
 and
  \begin{equation*}
    \big\| [q_u^{[j_1]} \otimes q_v^{[j_2]}, A^{\mathbf{1}}_t\, B^{\mathbf{1}}_t] \big\|_{\mathrm{op}} \;\leqslant\; \frac{4 n \|A^{\mathbf{1}}\|_{\mathrm{op}} \|B^{\mathbf{1}}\|_{\mathrm{op}}}{3 c N} (\text{e}^{\mathcal{V}t}-1)\,,\quad j_1,j_2\in\{n+1,\dots,n+m\}\,,
\end{equation*}
 whence,
 \begin{equation}\label{eq:I4}
  \begin{split}
   \bigg| \sum_{j_1=n+1}^{n+m}\sum_{j_2=n+1}^{n+m} \mathcal{E}_{j_1, j_2} \bigg|\;&\leqslant\; \| A^{\mathbf{2}}_t\|_{\mathrm{op}} \|B^{\mathbf{2}}_t\|_{\mathrm{op}} \sum_{j_1=n+1}^{n+m}\sum_{j_2=n+1}^{n+m} \big\| [q_u^{[j_1]} \otimes q_v^{[j_2]}, A^{\mathbf{1}}_t\, B^{\mathbf{1}}_t] \big\|_{\mathrm{op}} \\
   &\leqslant\;\frac{\,4m^2n\mathcal{L}}{3cN}(\text{e}^{\mathcal{V}t}-1)\,.
  \end{split}
 \end{equation}

  Combining \eqref{eq:I1}-\eqref{eq:I4} together we find
  \begin{equation}\label{eq:R-I}
   |\mathrm{(I)}|\;\leqslant\;\frac{\,4 n m \mathcal{L} }{3 c N}  (\text{e}^{ \mathcal{V}t}-1) \Big(n+m+ \frac{8 n m}{3 c N}(\text{e}^{\mathcal{V}t}-1) \Big)\,.
  \end{equation}

  Next, concerning now the first double sum in $\mathrm{(II)}$, the same procedure gives
  \[
   \mathcal{E}_{j_1, j_2}\;=\;\langle \,[A^{\mathbf{2}}_t\, B^{\mathbf{2}}_t, q_{v}^{[j_2]}] \,\,p_u^{[1, \dots, j_1-1]} \otimes p_v^{[1, \dots, j_2-1]}\,\,[q_u^{[j_1]}, A^{\mathbf{1}}_t\, B^{\mathbf{1}}_t]\, \rangle_{N_1, N_2}
  \]
  and, from \eqref{eq:ourLR},
  \[
   \begin{split}
     \big\| [q_u^{[j_1]}, A^{\mathbf{1}}_t\, B^{\mathbf{1}}_t] \big\|_{\mathrm{op}} \;&\leqslant\; \frac{4 n \|A^{\mathbf{1}}\|_{\mathrm{op}} \|B^{\mathbf{1}}\|_{\mathrm{op}}}{3 c N} (\text{e}^{\mathcal{V}t}-1)\,,\quad j_1\in\{n+m+1,\dots,N_1\}\,, \\
     \big\| [q_v^{[j_2]}, A^{\mathbf{2}}_t\, B^{\mathbf{2}}_t] \big\|_{\mathrm{op}} \;&\leqslant\; \frac{4 m \|A^{\mathbf{2}}\|_{\mathrm{op}} \|B^{\mathbf{2}}\|_{\mathrm{op}}}{3 c N} (\text{e}^{\mathcal{V}t}-1)\,,\!\quad j_2\in\{1,\dots,n\}\,,
   \end{split}
  \]
  whence
  \begin{equation}\label{eq:II1}
   \begin{split}
     \bigg| \sum_{j_1=n+m+1}^{N_1}\sum_{j_2=1}^{n} \mathcal{E}_{j_1, j_2} \bigg|\;&\leqslant\;  \sum_{j_1=n+m+1}^{N_1}\sum_{j_2=1}^{n}\big\| [q_u^{[j_1]}, A^{\mathbf{1}}_t\, B^{\mathbf{1}}_t] \big\|_{\mathrm{op}}\big\| [q_v^{[j_2]}, A^{\mathbf{2}}_t\, B^{\mathbf{2}}_t] \big\|_{\mathrm{op}} \\
     &\leqslant\;  \frac{\,16 (1-c)m n^2 \mathcal{L}}{9 c^2 N} (\text{e}^{\mathcal{V} t}-1)^2\,,
   \end{split}
  \end{equation}
  having used $\frac{N_1-m-n}{N}\leqslant 1-c$. \emph{Two} commutators were needed here in order to get a $O(N^{-1})$-smallness.

  In an analogous fashion, in the second double sum in $\mathrm{(II)}$ we insert commutators as
    \[
   \mathcal{E}_{j_1, j_2}\;=\; \langle \,[A^{\mathbf{2}}_t\, B^{\mathbf{2}}_t, q_u^{[j_1]}] \,\,p_u^{[1, \dots, j_1-1]} \otimes p_v^{[1, \dots, j_2-1]}\,\, [q_v^{[j_2]}, A^{\mathbf{1}}_t\, B^{\mathbf{1}}_t]\, \rangle_{N_1, N_2} 
  \]
  and we use \eqref{eq:ourLR} in the form
  \[
   \begin{split}
     \big\| [q_u^{[j_1]}, A^{\mathbf{2}}_t\, B^{\mathbf{2}}_t] \big\|_{\mathrm{op}} \;&\leqslant\; \frac{4 m \|A^{\mathbf{2}}\|_{\mathrm{op}} \|B^{\mathbf{2}}\|_{\mathrm{op}}}{3 c N} (\text{e}^{\mathcal{V}t}-1)\,,\quad j_1\in\{n+m+1,\dots,N_1\}\,, \\
     \big\| [q_v^{[j_2]}, A^{\mathbf{1}}_t\, B^{\mathbf{1}}_t] \big\|_{\mathrm{op}} \;&\leqslant\; \frac{4 n \|A^{\mathbf{1}}\|_{\mathrm{op}} \|B^{\mathbf{1}}\|_{\mathrm{op}}}{3 c N} (\text{e}^{\mathcal{V}t}-1)\,,\quad\; j_2\in\{n+1,\dots,n+m\}\,,
   \end{split}
  \]
   whence
  \begin{equation}\label{eq:II2}
   \begin{split}
     &\bigg| \sum_{j_1=n+m+1}^{N_1}\sum_{j_2=n+1}^{n+m} \mathcal{E}_{j_1, j_2} \bigg|\\
     &\qquad\leqslant\;  \sum_{j_1=n+m+1}^{N_1}\sum_{j_2=n+1}^{n+m}   \big\| [q_u^{[j_1]}, A^{\mathbf{2}}_t\, B^{\mathbf{2}}_t] \big\|_{\mathrm{op}} \big\| [q_v^{[j_2]}, A^{\mathbf{1}}_t\, B^{\mathbf{1}}_t] \big\|_{\mathrm{op}}  \\
     &\qquad\leqslant\;  \frac{\,16 (1-c)m^2 n \mathcal{L}}{9 c^2 N} (\text{e}^{\mathcal{V} t}-1)^2\,.
   \end{split}
  \end{equation}
  Again, \emph{two} commutators were crucial.

  From \eqref{eq:II1}-\eqref{eq:II2} we thus find
  \begin{equation}\label{eq:R-II}
   |\mathrm{(II)}|\;\leqslant\;\frac{\,16(1-c)m n(m+n)\mathcal{L}}{9 c^2 N} (\text{e}^{\mathcal{V}t} - 1)^2 \,.
  \end{equation}

  The contribution $\mathrm{(III)}$ has precisely the same form of $\mathrm{(II)}$ upon swapping A-particles and B-particles, thus under the exchange $j_1\leftrightarrow j_2$, $N_1\leftrightarrow N_2$, so we have at once
    \begin{equation}\label{eq:R-III}
   |\mathrm{(III)}|\;\leqslant\;\frac{\,16(1-c)m n(m+n)\mathcal{L}}{9 c^2 N} (\text{e}^{\mathcal{V}t} - 1)^2 \,.
  \end{equation}

  \subsection{Control of the terms of $O(N^2)$ size in $\mathcal{R}_{j_1,j_2}^{N_1,N_2}$}~

  The above strategy applied to the contribution $\mathrm{(IV)}$ of $\mathcal{R}_{j_1,j_2}^{N_1,N_2}$ leads apparently to a non-vanishing bound of order $O(1)$, because each of the $O(N^2)$-terms is shown, by insertion of \emph{two} commutators, to be $O(N^{-2})$-small. Actually, a more accurate algebraic estimate will fix this.

  Commutators are inserted in the usual manner in $\mathrm{(IV)}$ as
  \[
   \begin{split}
      \mathcal{E}_{j_1, j_2}\;&=\; \langle A^{\mathbf{2}}_t\, B^{\mathbf{2}}_t q_u^{[j_1]} \,\,p_u^{[1, \dots, j_1-1]} \otimes p_v^{[1, \dots, j_2-1]}  q_v^{[j_2]}\,\, q_u^{[j_1]}A^{\mathbf{1}}_t\, B^{\mathbf{1}}_t \rangle_{N_1, N_2} \\
      &=\;\langle \,[A^{\mathbf{2}}_t\, B^{\mathbf{2}}_t, q_u^{[j_1]}] \,\,p_u^{[1, \dots, j_1-1]} \otimes p_v^{[1, \dots, j_2-1]} q_v^{[j_2]}\,\, [q_u^{[j_1]},A^{\mathbf{1}}_t\, B^{\mathbf{1}}_t] \,\rangle_{N_1, N_2}\,,
   \end{split}
  \]
 having used $q_u^{[j_1]}= (q_u^{[j_1]})^2$ in the first identity, and \eqref{eq:ourLR} gives
   \[
   \begin{split}
     \big\| [q_u^{[j_1]}, A^{\mathbf{1}}_t\, B^{\mathbf{1}}_t] \big\|_{\mathrm{op}} \;&\leqslant\; \frac{4 n \|A^{\mathbf{1}}\|_{\mathrm{op}} \|B^{\mathbf{1}}\|_{\mathrm{op}}}{3 c N} (\text{e}^{\mathcal{V}t}-1)\,,\\
     \big\| [q_u^{[j_1]}, A^{\mathbf{2}}_t\, B^{\mathbf{2}}_t] \big\|_{\mathrm{op}} \;&\leqslant\; \frac{4 m \|A^{\mathbf{2}}\|_{\mathrm{op}} \|B^{\mathbf{2}}\|_{\mathrm{op}}}{3 c N} (\text{e}^{\mathcal{V}t}-1)\,,
   \end{split}\quad \quad j_1\in\{n+m+1,\dots,N_1\}\,.
  \]
  Exploiting these bounds and the linearity over $j_2$, which is an index \emph{not} appearing in the commutators, we find
  \begin{equation}\label{eq:to-ameliorate}
   \begin{split}
    &|\mathrm{(IV)}|\;\leqslant\;\bigg|\sum_{j_1 = n+m+1}^{N_1}\sum_{j_2=n+m+1}^{N_2}\mathcal{E}_{j_1, j_2}\bigg| \\
    &=\;\bigg|\sum_{j_1 = n+m+1}^{N_1}\bigg\langle \,[A^{\mathbf{2}}_t\, B^{\mathbf{2}}_t, q_u^{[j_1]}] \,\,p_u^{[1, \dots, j_1-1]}\otimes \bigg(\sum_{j_2=n+m+1}^{N_2} p_v^{[1, \dots, j_2-1]} q_v^{[j_2]} \bigg)\times \\
    &\qquad \qquad \qquad \qquad\qquad \qquad\times [q_u^{[j_1]}, A^{\mathbf{1}}_t\, B^{\mathbf{1}}_t]  \bigg\rangle_{N_1, N_2}\,\biggr| \\
    &\leqslant\;\bigg\|\sum_{j_2=n+m+1}^{N_2} p_v^{[1, \dots, j_2-1]} q_v^{[j_2]} \bigg\|_{\mathrm{op}}    
    \sum_{j_1 = n+m+1}^{N_1}\big\| [q_u^{[j_1]}, A^{\mathbf{1}}_t\, B^{\mathbf{1}}_t] \big\|_{\mathrm{op}} \big\| [q_u^{[j_1]}, A^{\mathbf{2}}_t\, B^{\mathbf{2}}_t] \big\|_{\mathrm{op}} \\
    &\leqslant\;\frac{\,16 \,m n\, \mathcal{L}}{9 c^2 N} (\text{e}^{\mathcal{V}t} - 1)^2\;\bigg\|\sum_{j_2=n+m+1}^{N_2} p_v^{[1, \dots, j_2-1]} q_v^{[j_2]} \bigg\|_{\mathrm{op}} \,.
   \end{split}
  \end{equation}

   The trivial estimate 
   \begin{equation*}
    \bigg\|\sum_{j_2=n+m+1}^{N_2} p_v^{[1, \dots, j_2-1]} q_v^{[j_2]} \bigg\|_{\mathrm{op}} \;\leqslant\;N
   \end{equation*}
   would make the bound \eqref{eq:to-ameliorate} ineffective. Instead, as an identity in the $N_2$-body Hilbert space, we re-write
   \[
    \begin{split}
     \sum_{j_2=n+m+1}^{N_2} p_v^{[1, \dots, j_2-1]} q_v^{[j_2]}\;&=\;\sum_{j_2=1}^{N_2} p_v^{[1, \dots, j_2-1]} q_v^{[j_2]}-\sum_{j_2=1}^{n+m} p_v^{[1, \dots, j_2-1]} q_v^{[j_2]} \\
     &=\;\mathbbm{1}_{N_2}-p_v^{\otimes N_2}-\sum_{j_2=1}^{n+m} p_v^{[1, \dots, j_2-1]} q_v^{[j_2]}
    \end{split}
   \]
   having applied the second formula from \eqref{identities2}. Using again the normalisation of orthogonal projections, this gives the ameliorated bound
    \begin{equation}\label{eq:goodbound}
    \bigg\|\sum_{j_2=n+m+1}^{N_2} p_v^{[1, \dots, j_2-1]} q_v^{[j_2]} \bigg\|_{\mathrm{op}} \;\leqslant\;2+n+m\,.
   \end{equation}
   Plugging \eqref{eq:goodbound} into \eqref{eq:to-ameliorate} then yields
   \begin{equation}\label{eq:R-IV}
    |\mathrm{(IV)}|\;\leqslant\;\frac{\,16(1-c)m n(2+n+m) \mathcal{L}}{9 c^2 N} (\text{e}^{\mathcal{V}t} - 1)^2\,.
   \end{equation}

  \subsection{Final estimate}~

  Combining \eqref{eq:Rsplit}, \eqref{eq:R-I}, \eqref{eq:R-II}, \eqref{eq:R-III}, and \eqref{eq:R-IV} together gives
  \begin{equation}\label{bounbj3}
   \begin{split}
      \big| \mathcal{R}_{j_1,j_2}^{N_1,N_2} \big| \; &\leqslant \;  \frac{\,4 n m \mathcal{L} }{3 c N}  (\text{e}^{ \mathcal{V}t}-1) \Big(n+m+ \frac{8 n m}{3 c N}(\text{e}^{\mathcal{V}t}-1) \Big)\\
        & \qquad \quad + \frac{\,32(1-c)m n(m+n)\, \mathcal{L}}{9 c^2 N} (\text{e}^{\mathcal{V}t} - 1)^2  \\
        & \qquad \quad +  \frac{\,16(1-c)m n(2+n+m) \mathcal{L}}{9 c^2 N} (\text{e}^{\mathcal{V}t} - 1)^2\,.
   \end{split}
  \end{equation}

  In turn, by means of \eqref{bounbj2}, \eqref{bounbj1}, and \eqref{bounbj3} the correlation function \eqref{sta} is estimated as
    \begin{equation}
   \begin{split}
    & \big|\,\big\langle A^{\mathbf{2}}_t  B^{\mathbf{2}}_t A^{\mathbf{1}}_t B^{\mathbf{1}}_t \big\rangle_{N_1, N_2}-\big\langle A^{\mathbf{2}}_t B^{\mathbf{2}}_t \big\rangle_{N_1, N_2} \big\langle A^{\mathbf{1}}_t B^{\mathbf{1}}_t \big\rangle_{N_1, N_2}\big| \\
    &\quad \leqslant\; \big|\mathcal{P}_{j_1}^{N_1}\big|+\big|\mathcal{Q}_{j_2}^{N_2}\big|+\big|\mathcal{R}_{j_1,j_2}^{N_1,N_2}\big| \\
    &\quad \leqslant\;\frac{\,4 m n \mathcal{L}}{3 c N}  (\text{e}^{\mathcal{V}t}-1) \Big(n+m+4+\frac{4(1-c)}{3 c}\Big(4+3m+3n+\frac{2nm}{N}\Big) (\text{e}^{\mathcal{V}t}-1)\Big)\,.
   \end{split}
  \end{equation}
   With the obvious bounds
   \[
     \text{e}^{\mathcal{V}t}-1\;\leqslant\;\text{e}^{2\mathcal{V}t}-1\,,\qquad (\text{e}^{\mathcal{V}t}-1)^2\;\leqslant\;\text{e}^{2\mathcal{V}t}-1
   \]
  last estimate becomes
  \begin{equation}
   \begin{split}
    & \big|\,\big\langle A^{\mathbf{2}}_t  B^{\mathbf{2}}_t A^{\mathbf{1}}_t B^{\mathbf{1}}_t \big\rangle_{N_1, N_2}-\big\langle A^{\mathbf{2}}_t B^{\mathbf{2}}_t \big\rangle_{N_1, N_2} \big\langle A^{\mathbf{1}}_t B^{\mathbf{1}}_t \big\rangle_{N_1, N_2}\big| \\
    &\quad \leqslant\;\frac{\,4 m n \mathcal{L}}{9 c^2 N}(\text{e}^{2\mathcal{V}t}-1) \Big( \frac{8 m n (1-c)}{N}+4(4-c)+3(4-3c)(m+n) \Big) \\
    &\quad \leqslant\;\frac{\,4 m n \mathcal{L}}{9 c^2 N}(\text{e}^{2\mathcal{V}t}-1) \Big( \frac{8 m n}{N}+4(4+3m+3n)\Big)
   \end{split}
  \end{equation}
  which has precisely the form \eqref{claimthgrowth2}.

  This completes the proof of Theorem \ref{thm:main1}.

%
%
%

  \def\cprime{$'$}

\end{document}